\documentclass[11pt]{article}

\usepackage{amssymb}
\usepackage{amsfonts}
\usepackage{amsmath}
\usepackage{cite}
\usepackage{mathrsfs}
\usepackage{graphicx}%
\usepackage{verbatim}
\usepackage{color}
\usepackage{makecell} 
\usepackage{multirow}
\usepackage{enumitem}
\usepackage{algorithm}
\usepackage{algpseudocode}
\usepackage{subeqnarray}
\usepackage{amsmath,bm}
\usepackage{subfig}
\usepackage{array,calc}
\usepackage{bbm}
\usepackage{hyperref}
\usepackage{dsfont}
\usepackage{authblk}
\usepackage[title]{appendix}
\newcommand{\RNum}[1]{\uppercase\expandafter{\romannumeral #1\relax}}
\makeatother

\allowdisplaybreaks[4]
\usepackage{algorithm}
\usepackage{algpseudocode}
\usepackage{setspace}
\algnewcommand{\Inputs}[1]{%
	\State \textbf{Inputs:}
	\Statex \hspace*{\algorithmicindent}\parbox[t]{.9\linewidth}{\raggedright #1}
}
\algnewcommand{\Initialize}[1]{%
	\State \textbf{Initialization:}
	\Statex \hspace*{\algorithmicindent}\parbox[t]{.9\linewidth}{\raggedright #1}
}

\providecommand{\U}[1]{\protect\rule{.1in}{.1in}}
\begin{document}
	
\title{Towards the efficacy of federated prediction for epidemics on networks}
\author[1,2]{Chengpeng Fu}
\author[2]{Tong Li}
\author[1]{Hao Chen}
\author[1,2,3]{Wen Du}
\author[1,2]{Zhidong He\thanks{zhidonghe@outlook.com}}

\affil[1]{DS Information Technology Co., Ltd., Shanghai, China.}
\affil[2]{The First Research Institute of Telecommunications Technology Co., Ltd., Shanghai, China.}
\affil[3]{Wuhan Yangtze Communications Industry Group Co., Ltd., Wuhan, China.}
\date{}
\setcounter{Maxaffil}{0}
\renewcommand\Affilfont{\itshape\small}
\maketitle

\begin{abstract}
Epidemic prediction is of practical significance in public health, enabling early intervention, resource allocation, and strategic planning. However, privacy concerns often hinder the sharing of health data among institutions, limiting the development of accurate prediction models. 
In this paper, we develop a general privacy-preserving framework for node-level epidemic prediction on networks based on federated learning (FL). We frame the spatio-temporal spread of epidemics across multiple data-isolated subnetworks, where each node state represents the aggregate epidemic severity within a community. Then, both the pure temporal LSTM model and the spatio-temporal model i.e., Spatio-Temporal Graph Attention Network (STGAT) are proposed to address the federated epidemic prediction.
Extensive experiments are conducted on various epidemic processes using a practical airline network, offering a comprehensive assessment of FL efficacy under diverse scenarios.
By introducing the efficacy energy metric to measure system robustness under various client configurations, we systematically explore key factors influencing FL performance, including client numbers, aggregation strategies, graph partitioning, missing infectious reports.
Numerical results manifest that STGAT excels in capturing spatio-temporal dependencies in dynamic processes whereas LSTM performs well in simpler pattern. Moreover, our findings highlight the importance of balancing feature consistency and volume uniformity among clients, as well as the prediction dilemma between information richness and intrinsic stochasticity of dynamic processes.
This study offers practical insights into the efficacy of FL scenario in epidemic management, demonstrates the potential of FL to address broader collective dynamics.
\end{abstract}

\section{Introduction}
Epidemic modeling has a long and significant history, dating back to Daniel Bernoulli’s work in 1766 \cite{bernoulli1760essai}, which laid the foundation for understanding the dynamics of disease spreading. Epidemics on networks represent a ubiquitous process, capturing diverse phenomena such as infectious disease transmission \cite{pastor2015epidemic}, neural excitation, cascading failures in networked systems \cite{smolyak2020mitigation}, and the dissemination of information or rumors \cite{vosoughi2018spread}.
The global COVID-19 pandemic underscored the critical need for effective epidemic prediction \cite{singh2020prediction}. The ability to anticipate outbreaks facilitates early warnings, enabling the establishment of control measures, quarantine planning, and optimal allocation of medical resources \cite{he2018optimal}, which are vital for minimizing damage during outbreaks.

Epidemic prediction can be framed as a spatio-temporal prediction problem due to the underlying contagion process driven by contact interactions between individuals or regions. 
Predicting node states over a future period of time, as a fine-grained approach to epidemic modeling, naturally aligns with spatio-temporal classification tasks.
Temporal dependencies are effectively captured using data-driven methods like Long Short-Term Memory (LSTM) networks, which have demonstrated substantial success across various domains \cite{van2020review}. 
Meanwhile, Graph Neural Networks (GNNs) and its temporal variants, have emerged as powerful tools for learning the complex topological relationships within networked populations \cite{wu2020comprehensive}. These methods have been successfully applied in fields such as recommendation \cite{chen2020tgcn}, traffic prediction \cite{li2021spatial}, and drug discovery \cite{jimenez2020drug}, which provides a robust basis for spatio-temporal modeling.

In the context of a global pandemic, governments and health organizations in various districts or countries often collect sensitive individual health data to monitor and manage outbreaks within their jurisdictions. However, there is a notable reluctance among these entities to publicly share or pool their data, largely due to privacy concerns, data ownership, and regulatory constraints.
This presents a critical challenge: how can health centers and governments be persuaded to collaborate in a privacy-preserving manner while still enabling effective epidemiological prediction?
Federated learning (FL) offers a promising solution to this dilemma by facilitating collaborative machine learning across distributed data sources without the need for centralized data sharing \cite{konevcny2016federated}. 
Federated learning has been successfully applied in keyboard prediction, knowledge graph construction \cite{wu2023federated}, and other domains. 
By enabling the construction of robust, global-scale epidemiological models that rely on locally-sourced data, FL empowers health authorities to make more informed decisions without sacrificing privacy, thus fostering a more cooperative and efficient response to future pandemics.
A more detailed review of related works is presented in Section 5.

Previous studies on federated frameworks for epidemics remain limited and overly specific, often focusing on narrow scenarios or single epidemic models \cite{samuel2022iomt,fu2024privacy,kumaresan2022analysis}. 
Instead of pursuing a SOTA method for a specific dataset, this paper aims to propose a general federated framework for epidemic prediction on networks, complemented by a systematic approach to assess the efficacy of federated learning from a statistical perspective.
We model the transmission dynamics of epidemics within confined districts, where a network $G$ with $N$ nodes is partitioned into $M$ regions, each representing a distinct client. Each client retains its local subnetwork and epidemic data, while a central server, managed by an authoritative organization such as the CDC or WHO, oversees and coordinates the federated learning process.
The proposed framework accommodates diverse epidemic processes, spanning from two-state models like SIS to multi-state models like SIRVS, encompassing both Markovian and non-Markovian dynamics, as well as static and seasonal infection rates.
Beyond leveraging LSTM for temporal modeling, we propose a novel Spatio-Temporal Graph Attention Network (STGAT) for federated epidemic prediction. STGAT incorporates a spatial dependency extraction block, based on Graph Attention Networks (GAT), alongside a temporal feature extraction block, effectively capturing complex epidemic dynamics \cite{huang2019stgat}. The attention mechanism in GAT facilitates the computation of pairwise correlations between nodes, enabling the model to exploit intricate interactions within the network. 

To evaluate our proposed framework, we conduct comprehensive experiments on a practical airline network. Unlike previous studies that focus on predetermined scenarios with fixed client configurations, our evaluation systematically considers multiple factors influencing the overall performance of the FL system. These factors include performance metrics, client configurations, prediction models, aggregation methods, epidemic processes, and network partitions strategies and missing reports. 
Additionally, we introduce a novel performance indicator, referred to as ``efficacy energy'', defined as the average performance across an increasing number of clients, which provides a robust measure of the effectiveness and scalability of federated learning scenarios, capturing the system's resilience under dynamic client configurations. 
One can refer to our project\footnote{\url{https://github.com/S1mple-yyds/Fed_Epi_Classifier}} on Github for the related source. The main contributions of this work can be summarized as follows:

1) We formulate a framework for addressing the federated epidemic prediction problem, utilizing both LSTM and STGAT method. Our results demonstrates that STGAT does not consistently outperform LSTM across all scenarios.

2) We introduce a comprehensive evaluation methodology for federated epidemic prediction, incorporating detailed considerations of factors influencing performance. Additionally, the metric of efficacy energy is proposed to provide a holistic measure of system robustness under dynamic conditions.

3) Through extensive experiments on a real-world airline network, we systematically evaluate the proposed framework across diverse scenarios. The findings highlight the significance of an elaborate setting for federated epidemic prediction, and offer valuable insights to guide future optimizations.

This paper is organized as follows. Section 2 introduce preliminary knowledge about epidemics and defines the problem. The federated framework for epidemic prediction is proposed in Section 3. We then conduct experiments and discussion the findings in Section 4. Last, we introduce the related work in Section 5 and conclude this paper in Section 6.

\section{Preliminaries of Epidemics on networks}
In this section, we provide an overview of our fundamental concepts and formally propose the problem of epidemic prediction on networks. A brief introduction of compartmental models in epidemiology is introduced as follows.

\subsection{Epidemics on networks}
The concerned dynamics in this paper is \textit{epidemic}, also called non-conserved spread \cite{newman2006structure}, where the individuals are divided into several compartments, e.g., the susceptible, the infectious, or the recovered. The nodal states switch among compartments both by the contacts between individuals and by spontaneous processes.
Previous researches have provided an exhaustive investigation on the phase transition behaviors and estimation of prevalence for different epidemic models \cite{pastor2015epidemic} \cite{pastor2001epidemic}.

Considering the simplest Susceptible-Infected-Susceptible (SIS) on networks as an representative process, each infected item can be cured, and becomes susceptible again after recovering from the infection state. Both the curing and infection processes are Poisson processes in Markovain SIS epidemics \cite{van2014performance}. 
Specifically, we define a Bernoulli random variable $X_i(t)\in\{0,1\}$ as the state of a node $i$ at time $t$, where $X_i(t)=0$ for the susceptible state and $X_i(t)=1$ for the infected state. The network $G$ with $N$ nodes and $L$ links is represented by an adjacent matrix $A$, where $a_{ij}=1$ if there is a link between node $i$ and node $j$, otherwise $a_{ij}=0$. We denote by $\mathcal{N}=\{1,2,\dots,N\}$ the set of nodes in the network. 
Following the nodal infection probability $E[X_i]=\Pr[X_i=1]$, the exact SIS governing equation for node $i$ follows
\begin{equation}
	\frac{dE[X_i(t)]}{dt}=E\left[-\delta X_i(t)+\beta(1-X_i(t))\sum_{k=1}^N a_{ki}X_k(t)\right]	
\end{equation} 

The ratio between the infection rate $\beta$ and the curing rate $\delta$ is called the effective infection rate $\tau=\beta/\delta$. 
The SIS model features a phase transition \cite{castellano2010thresholds} around the epidemic threshold $\tau_c$. Viruses with an effective infection rate $\tau$ above the epidemic threshold $\tau_c$ can infect a sizeable portion of the population on average and stay for a long time in the network. This long period is called the metastable state. 
A first-order mean-field approximation of the epidemic threshold $\tau_c^{(1)}= 1/ \lambda_1(A)$, where $\lambda_1(A)$ is the spectral radius of the adjacency matrix $A$, was shown \cite{van2013non} to be a lower bound for the epidemic threshold, $\tau_c^{(1)}<\tau_c$.

\begin{figure}[htbp]
	\centering
	\begin{minipage}[b]{0.55\textwidth}
		\centering
		\includegraphics[width=\textwidth]{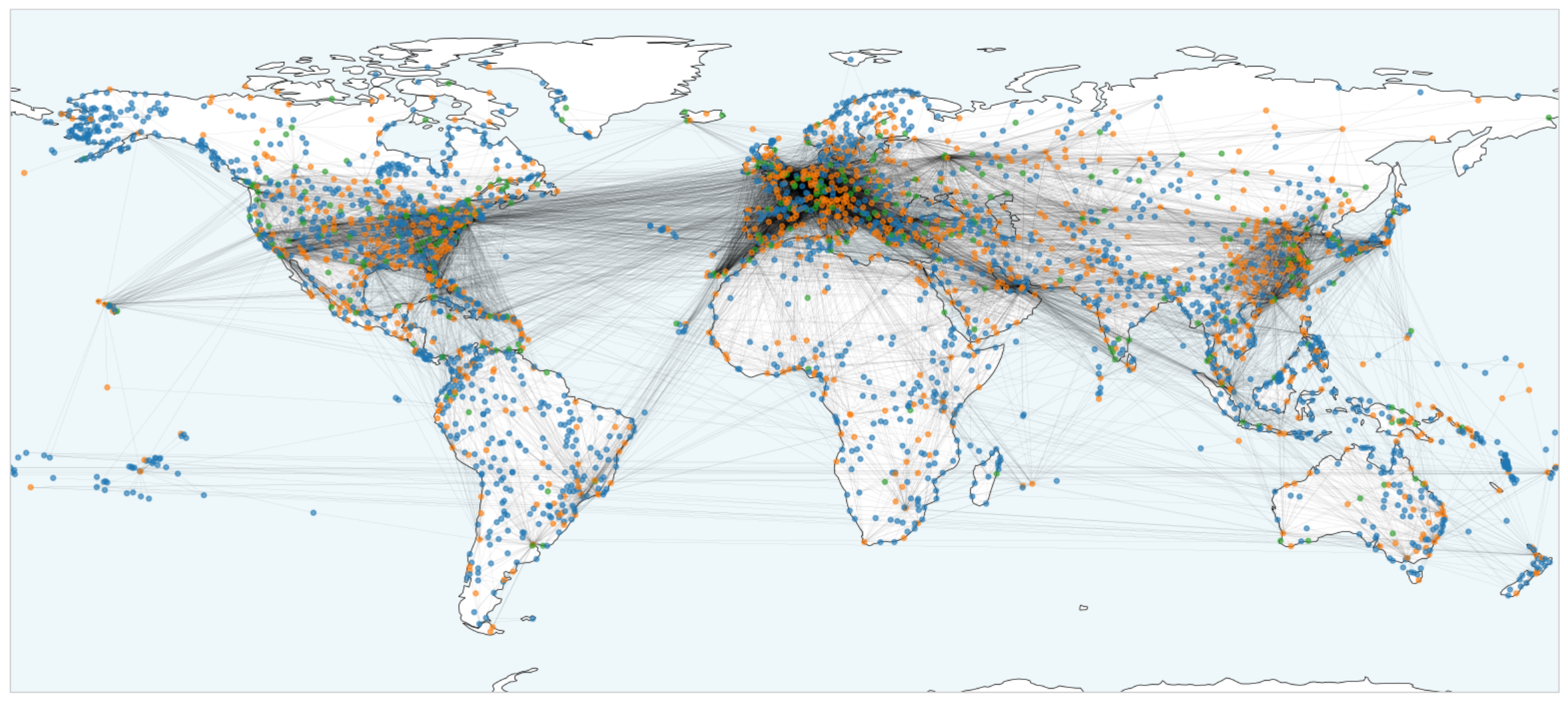} 
	\end{minipage}
	\hspace{0.5cm}
	\begin{minipage}[b]{0.2\textwidth}
		\centering
		\includegraphics[width=0.95\textwidth]{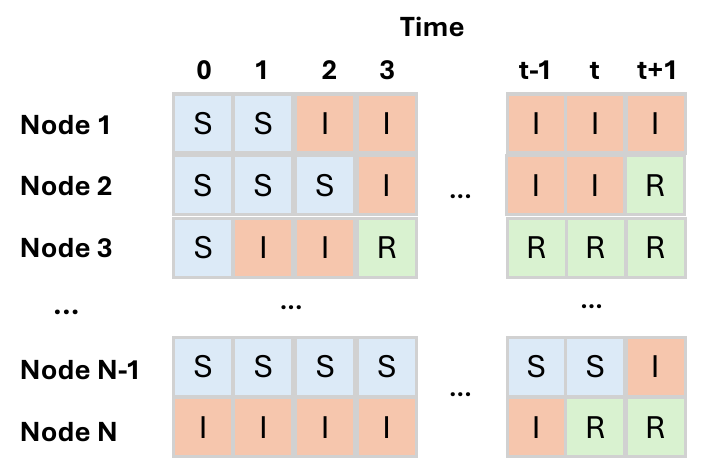} 
		\vspace{-0.1cm} 
		\includegraphics[width=\textwidth]{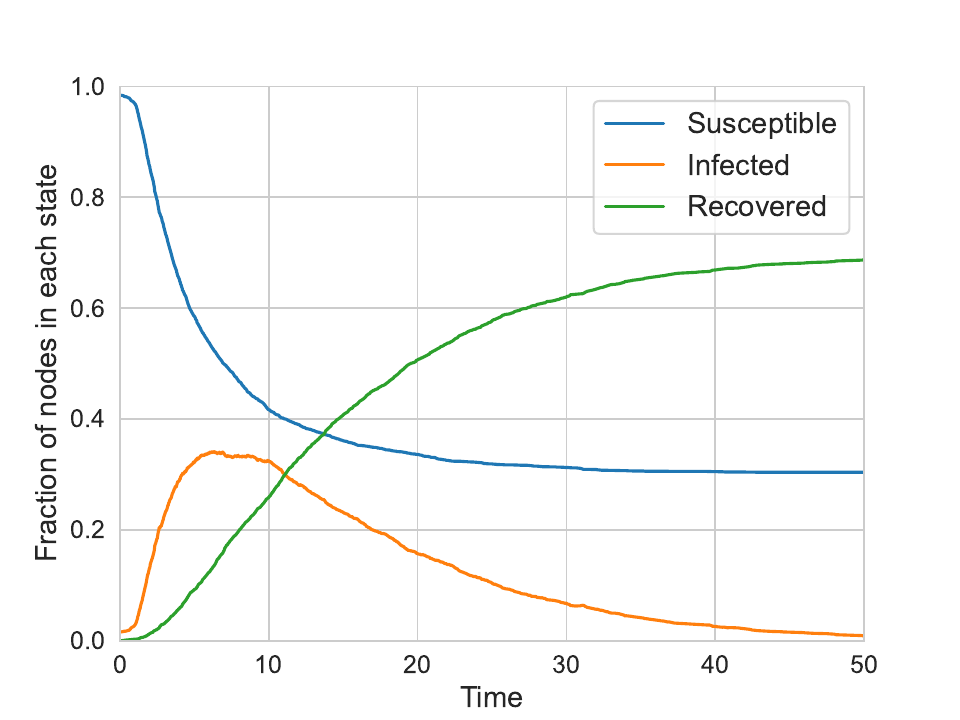} 
	\end{minipage}
	\caption{Illustration of an SIR spreading on a network. The left subfigure shows a snapshot at $t=5$, where the blue nodes represent the susceptible, the orange represents the infected, and the greed for the recovered. The upper right subfigure exemplifies temporal node states, and the bottom right subfigure present the prevalence of this SIR spreading.}
	\label{fig:epi_models}
\end{figure}

We usually assessing the severity and the tendency of an epidemic via investigating the time-varying fraction of infected nodes, which is defined as the prevalence, i.e., $y_I(t) = \frac{1}{N}\sum_{i=1}^N\bm{1}_{X_i(t)=1}$ where $\bm{1}_{\{X_i(t)=1\}}$ is an indicator function that takes the value 1 if node $i$ is in the infection state $I$ at time $t$, and 0 otherwise.
The exact Markovian SIS model \cite{van2009virus} in the network $G$ with $N$ nodes consists of $2^N$ states, which is intractable to solve and prediction analytically for large networks. 
In this paper, we investigate the feasibility of federated graph learning for more extended compartmental models in epidemiology, which refer to Table ~\ref{tab:epi_models}. 
The considered processes encompass a range from two-state models such as SIS to multi-state models like SIRVS, incorporating both Markovian and non-Markovian dynamics, along with infection rates that may be static or time-varying.
Due to the limitation of the analytical methods, an event-driven simulator for the spreading process based on the Gillespie algorithm is implemented to simulate the spatio-temporal process of epidemic spread on networks \cite{he2018spreading}. Figure  ~\ref{fig:epi_models} illustrates an example of SIR epidemic on a network.

\begin{table}[tb]
		\centering
		\footnotesize
		\renewcommand{\arraystretch}{1.4}
		\scalebox{0.85}{
			\begin{tabular}{p{1in} | p{4in} | p{2.2in}}
				\hline
				\hline						
				Model &  Description & Factors \tabularnewline			
				\hline				
				SIS\cite{pastor2015epidemic} &  The spread where individuals can transition between susceptible (S) and infected (I) states, with infected individuals recovering and becoming susceptible again & \makecell[tl]{$\beta$ - the infection rate\\$\delta$ - the curing rate}     \\
				SIR\cite{pastor2015epidemic} &  The spread where individuals transition from susceptible (S) to infected (I) and then to recovered (R), with recovered individuals gaining immunity and no longer participating in the infection dynamics & \makecell[tl]{$\beta$ - the infection rate\\$\delta$ - the curing rate}     \\
				SEIR\cite{pastor2015epidemic} &  The spread of a disease where individuals progress through four states: susceptible (S), exposed (E, infected but not yet infectious), infected (I), and recovered (R), with recovered individuals gaining immunity. & \makecell[tl]{$\beta_1$ - the infection rate\\$\beta_2$ - the incubation rate\\$\delta$ - the curing rate}     \\
				non-Markovian SIS\cite{van2013non} & The non-Markovian SIS epidemic model modifies the standard SIS dynamics by using a Weibull distribution (instead of an exponential distribution) for the transmission time, allowing for more flexible and realistic infection and recovery times.  & \makecell[tl]{$\beta_\lambda$-the scale parameter \\$\beta_k$-the shape parameter\\$\delta$-the curing rate}     \\
				SIRS\cite{pastor2015epidemic} &  The dynamics where individuals transition from susceptible (S) to infected (I), then to recovered (R), but after a period of immunity, they can become susceptible again. & \makecell[tl]{$\beta$ - the infection rate\\$\delta$ - the curing rate\\$\omega$ - the immunity loss rate}      \\
				SIRVS\cite{zhang2008sirvs} &  
				The SIRVS epidemic model extends the SIR framework by incorporating vaccinated (V) and susceptible (S) states, where individuals can transition from susceptible to vaccinated, potentially reducing infection rates and influencing the disease dynamics. & \makecell[tl]{$\beta$ - the infection rate\\$\delta$ - the curing rate\\$\omega$ - the immunity loss rate\\$v_1$ - the vaccination rate\\$v_2$ - the vaccine waning rate }   \\
				SIS with time-varying rates \cite{crokidakis2012critical} &  
				The SIS epidemic model with time-dependent rates incorporates a seasonal variation in the infection rate, defined as the infection rate $\beta(t)=a+b\sin(t/c)$ in this paper,  where $a,b,c$ are parameters.
				& \makecell[tl]{$\beta(t)$ - the infection rate \\$\delta$ - the curing rate }      						
				\tabularnewline
				\hline
				\hline
			\end{tabular}
		}
		\caption{Summary of compartmental epidemic models under investigation in this paper.}
		\label{tab:epi_models}
\end{table}

\subsection{Problem of epidemic prediction}
The propagation of infectious diseases is fundamentally driven by interactions between individuals, with such interactions often represented as a contact network $G$. Accurately predicting the infection status of individuals within this network is critical for the effective design of targeted intervention strategies \cite{he2018optimal}. However, in practical scenarios, the precise model of disease transmission is often unknown and based on assumptions. 
Moreover, due to the stochastic nature of disease spread, parameters such as infection rate, incubation period, and recovery rate are inherently uncertain, which underscores the motivation for data-driven approaches in transmission forecasting. Regarding that the state of each node represents either the status of an individual or the macroscopic state of a community \cite{moon2020group}, the epidemic prediction problem on a network can be framed as a temporal node classification task, according to the historical status of nodes and the topology of contact relations.

\textbf{Definition 1: Epidemic Prediction Problem.} Given an un-directed contact network $G=(N,L)$ with $N$ nodes and $L$ links, the goal of epidemic prediction is to predict the future epidemic transmission over the upcoming $t_F$ time steps based on historical status trajectory of all individuals from the past $t_H$ time steps, through a trained model $f(\cdot)$.
\begin{align}
	\hat{\mathbf{X}}_{t+1:t+t_F}= \hat{f}\left( \mathbf{X}_{t-t_H+1:t}; G,\Theta \right)
\end{align}
where $x_i(t)$ represents the state of node $i$ at time $t$, and $x_i(t)\in\{S,I,R,...\}$, and $\Theta$ is parameters of model. The state vector $\mathbf{x}_t = \{x_1(t), x_2(t), \dots, x_N(t)\}$ indicates the state of all nodes at time $t$, and the state set $\mathbf{X}_{t_1:t_2} = \{\mathbf{x}_{t_1}, \mathbf{x}_{t_1+1}, \dots, \mathbf{x}_{t_2}\}$ comprises the state of all nodes from time $t_1$ to $t_2$.

\section{Federated framework for epidemic prediction}
In this section, we provide the framework of our federate leaning system and formulate the problem of federated epidemic prediction. We also introduce the fed-LSTM model and fed-STGAT model for the spatio-temporal dynamics forecasting with privacy-preserving. 

\subsection{System Overview and problem statements}
\begin{figure}[tb]\centering 
	\includegraphics[width=12cm]{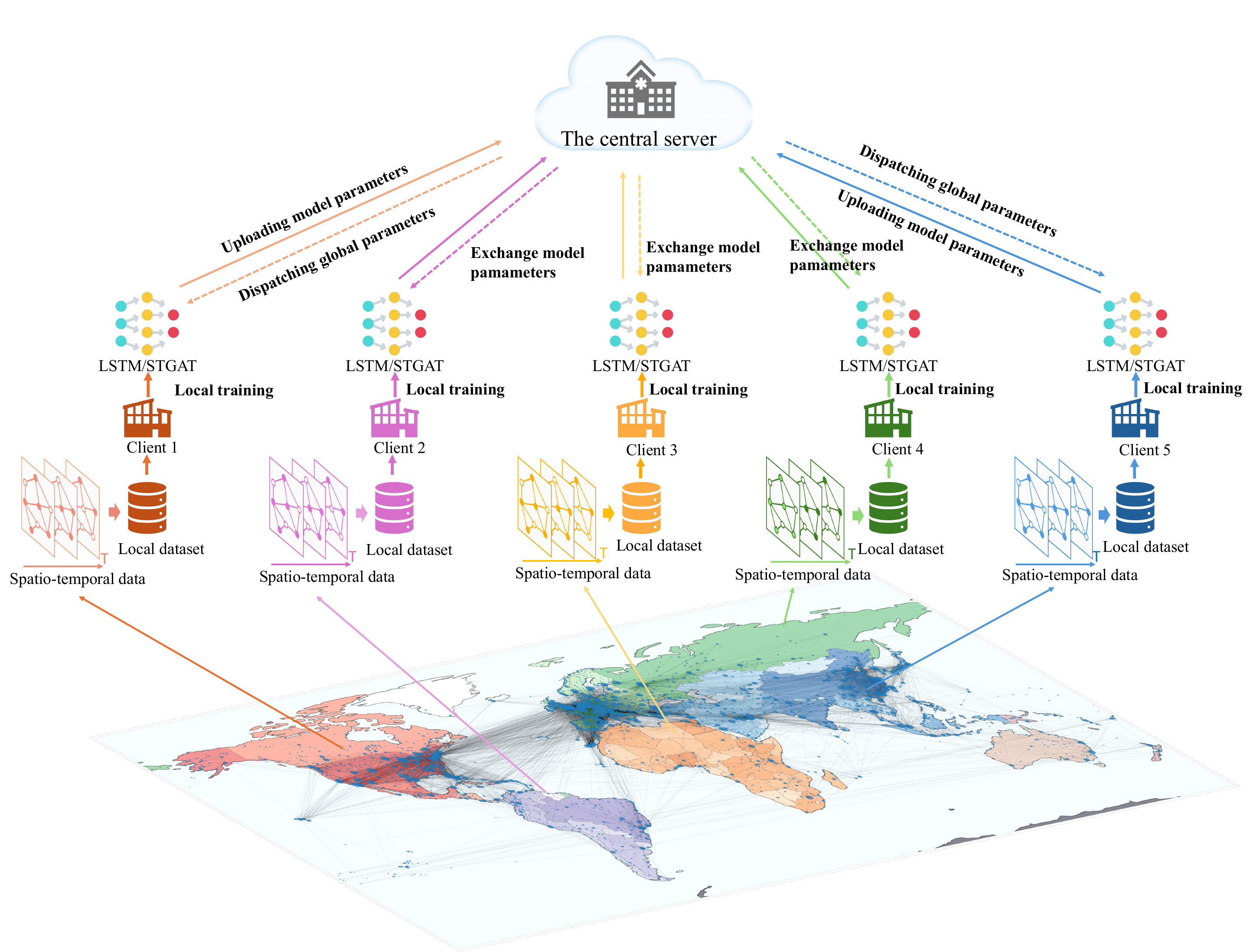}
	\caption{Illustration of the proposed federated learning framework designed to epidemic prediction. A central server coordinates the training process, while each client performs effective local training based on its own temporal node states and subnetwork topology.}
	\label{fig:framework}
\end{figure}

Real-world epidemic prediction faces challenges arising from data availability and privacy concerns. Early-stage data collection is often sparse, and privacy concerns restrict the sharing of sensitive health data among governments and health organizations, which hinders the development of accurate prediction models. Federated learning enables decentralized training of machine learning models across distributed data while preserving privacy.
As illustrated in Figure \ref{fig:framework}, the proposed federated learning framework for epidemic prediction operates within a complex network $G$ comprising $N$ individuals. Epidemics frequently transcend regional and national borders, escalating to a global scale. Within this context, the state of an individual node in the model represents not only the health status of an individual but also serves as an aggregate indicator of the epidemic’s severity within a community, city, or district \cite{moon2020group}\cite{bonaccorsi2015epidemic}. This abstraction enables individual node states to be interpreted as proxies for community-level health metrics, offering a comprehensive perspective on the epidemic’s impact across varying spatial and social scales \cite{del2023epidemic}.

The FL system consists of $M$ regions, with each region comprising a subset of nodes and representing a distinct client. Usually, certain regions may share nodes, reflecting overlapping boundaries or inter-regional interactions. Each client’s local device retains the infection prediction results and historical trajectories of its nodes, ensuring that data remains decentralized. For instance, a sentinel hospital within each subnetwork can monitor infection dynamics over time, facilitating localized data collection and processing.
A central server, hosted by an authoritative organization such as a government entity (e.g., the CDC) or an international agency (e.g., the WHO), oversees and coordinates the federated learning process. This decentralized architecture enables the aggregation of information across regions without directly accessing raw individual or regional data, thereby preserving privacy while ensuring robust and comprehensive epidemic predictions.
We denote $G^{(m)}$ the subnetwork of the $m$-th region for $m\in\{1,2,\dots,M\}$, which consist of $N^{(m)}$ nodes. The problem of federated epidemic prediction is defined as follow:

\textbf{Definition 2: Federated Epidemic Prediction Problem (FEP).} 
Let $G$ be a contact network comprising $N$ nodes, partitioned into $M$ disjoint or overlapping subnetworks $\{G^{(1)}, G^{(2)}, \dots, G^{(M)} \}$, where each subnetwork $G^{(m)}\subseteq G$ contains $N^{(m)}$ nodes for $m \in \{1, 2, \dots, M\}$. The objective of federated epidemic prediction is to forecast the future states of the nodes in the network for the upcoming $t_F$ time steps, based on the historical trajectory of node states in each subnetwork over the preceding $t_H$ time steps. Formally, this can be expressed as learning a predictive function $f : \mathcal{X}_H \times \mathcal{G} \rightarrow \mathcal{X}_F$, where
\begin{itemize}
	\item $\mathcal{X}_H = \{ \mathbf{X}_{t-H+1:t}^{(m)} \mid m = 1, \dots, M \}$ represents the historical node states in each subnetwork over the past $t_H$ time steps,
	\item $\mathcal{G} = \{ G^{(m)} \mid m = 1, \dots, M \}$ denotes the set of subnetwork capturing the topological structure of each region,
	\item $\mathcal{X}_F = \{ \hat{\mathbf{X}}_{t+1:t+t_F}^{(m)} \mid m = 1, \dots, M \}$ represents the predicted future node states over the next $t_F$ time steps.
\end{itemize}


\subsection{Federated Learning Process}
In the architecture of federated learning (FL), the training process is distributed across multiple clients (e.g., regions, hospitals, or devices), and a central server orchestrates the global model updates. The following outlines the core steps of server-side processing and client-side training, where two federated learning approaches, i.e., FedAvg \cite{mcmahan2017communication}, FedProx \cite{li2020federated}, are applied in our study.

The server is responsible for maintaining and updating the global model parameters, denoted as $\bm{\theta}$. At the beginning of each training round, the server distributes the current global model $\bm{\theta}$ to all participating clients. Each client then performs local training based on its local data and returns the computed gradients or updated model parameters to the server. The server aggregates the received updates and computes a new set of global model parameters.
For the FedAvg approach, the server aggregates the model updates from clients using a weighted average, where the weight corresponds to the number of data samples on each client. The global model update for FedAvg can be expressed as:
\begin{align}
\bm{\theta} = \sum_{i=1}^{M} \frac{n_i}{N} \bm{\theta}_i
\end{align}
where $\bm{\theta}_i$ is the local model from the $i$-th client, $n_i$ is the number of data samples on the $i$-th client, and $N$ is the total number of data samples across all clients. Each client $i$ receives the global model parameters $\theta$ from the server at the start of each round. Using its local dataset, the client performs local training for multiple epochs. For FedAvg, the local model on each client is updated by minimizing the local objective function:
\begin{align}\label{equ:avg}
\bm{\theta}_i' = \arg \min_{\bm{\theta}_i} \left( \ell_i(\bm{\theta})\right)
\end{align}
where $\ell_i(\bm{\theta})$ is the loss function on client $i$. After training, the client sends either the updated local model $\bm{\theta}_i'$. 
For FedProx, the server-side aggregation is similar to FedAvg, but the local training objective includes a proximal term that discourages significant deviation from the global model. The local model on each client is updated by minimizing the local objective function:
\begin{align}\label{equ:prox}
\bm{\theta}_i' = \arg \min_{\bm{\theta}_i} \left( \ell_i(\bm{\theta}) + \frac{\mu}{2} \|\bm{\theta} - \bm{\theta}_i\|^2 \right)
\end{align}
where $\mu$ is the proximal term in FedProx that helps stabilize local updates when data distributions across clients are heterogeneous.
 The overall process can be summarized by Algorithm \ref{alg:fep}.
\begin{algorithm}[H]
	\caption{Federated Learning Process for Epidemic Prediction (FEP)}
	\begin{algorithmic}[1]
		\State \textbf{Server Initialization:} Initialize global model $\bm{\theta}$.
		\For {each round}
		\State Server sends global model $\bm{\theta}$ to all clients.
		\For {each client $i$}
		\State Client receives global model $\bm{\theta}$ from server.
        \If {method == FedAvg}
							\State \makebox[3.5cm][l]{Client updates local model $\bm{\theta}_i$ by formula (\ref{equ:avg})}
		\ElsIf {method == FedProx}
					\State \makebox[3.5cm][l]{Client updates local model $\bm{\theta}_i$ by formula (\ref{equ:prox})}
                    
        \EndIf
		\State Client sends local update $\bm{\theta}_i'$ to the server.
		\EndFor
		\State Server aggregates updates: $\bm{\theta} = \sum_{i=1}^{M} \frac{n_i}{N} \bm{\theta}_i'$
		\State Server updates global model $\bm{\theta}$
		\If {Early Stop condition satisfried}
		\State Break the loops.
		\EndIf
		\EndFor
	\end{algorithmic}
	\label{alg:fep}
\end{algorithm}

\subsection{LSTM for epidemic prediction}
In this study, we leverage the Long Short-Term Memory (LSTM) network for epidemic prediction of time-series classification, where the input consists of the state of all nodes in a network over the past $t_H$ time steps, and the output is the predicted state of all nodes for the upcoming $t_F$ time steps. The LSTM model is specifically designed to capture temporal dependencies in sequential data, making it suitable for modeling the progression of an epidemic over time based on historical node states. The architecture of our LSTM model comprises several key components, described as follows:

\textbf{Embedding Layer:} 
The primary input to our model is the sequence of node states over the time window $[t_0 - t_H + 1, t_0]$. For each time step $t$ in this window, the state of all nodes is denoted by the vector $\mathbf{x}_{t} = \{x_1(t), x_2(t), \dots, x_N(t)\}$, where each node's state $x_i(t)$ belongs to one of multiple classes (e.g., susceptible, infected, recovered). To transform these categorical states into a continuous vector space, we introduce an embedding layer that maps each node's state $x_i(t)$ into a dense vector representation $e_i(t) \in \mathbb{R}^d$, where $d$ is the embedding dimension. 

\textbf{LSTM Layer:} The core of the architecture is the LSTM layer, which is responsible for learning temporal dependencies in the sequence of node states. At each time step $t$, the LSTM cell takes as input the embedded state vector $e_i(t)$ and the hidden state from the previous time step $h_{t-1}$, and outputs a new hidden state $h_t$. The LSTM cell consists of three key gates—input gate, forget gate, and output gate—that regulate the flow of information into and out of the memory cell, enabling the model to retain or discard information from past time steps as needed. 

Forget Gate determines which past information should be discarded, which follows
\begin{align}
	f_t = \sigma \left( W_f [h_{t-1}, e_t] + b_f \right)
\end{align}
where $h_{t-1}$ is the hidden state from the previous time step, and $e_t$ is the input from the embedding layer.
Input Gate decides which new information to store in the cell state and the update of the state, which follows
\begin{align}
	i_t &= \sigma \left( W_i [h_{t-1}, e_t] + b_i \right)\\
	\tilde{c}_t &= \tanh \left( W_C [h_{t-1}, e_t] + b_c \right)\\
	c_t &= f_t \cdot c_{t-1} + i_t \cdot \tilde{c}_t
\end{align}
where $c_t$ is the updated cell state at the current time step.
\textit{Input Gate} produces the hidden state $h_t$ that will be passed to the next LSTM unit, which follows
\begin{align}
	o_t &= \sigma(W_o \cdot [h_{t-1}, e_t] + b_o)\\
	h_t &= o_t \cdot \tanh(c_t)
\end{align}

\textbf{Output Layer:}
The hidden state $h_t$ from the final LSTM cell is passed through a softmax layer, which converts the output into a probability distribution over the possible classes for each node's state (e.g., susceptible, infected, recovered). The softmax function is given by:
\begin{align}\label{equ:softmax}
	p_t = \text{softmax}(W_y h_t + b_y)
\end{align}
where $p_t$ represents the probability distribution over the node states. The predicted states for the nodes over the interval $[t+1, t+t_F]$ are computed by selecting the class with the highest probability for each node:
\begin{align}\label{equ:argmax}
	\hat{\mathbf{X}}_{t+1:t+t_F} = \arg \max(p_t)
\end{align}

Given the historical states of all nodes $\mathbf{X}_{t-t_H+1:t}$, the LSTM processes the sequence and generates predictions for the future states $\hat{\mathbf{X}}_{t+1:t+t_F}$. The model is trained to minimize the cross-entropy loss between the predicted and actual node states over the future time steps, allowing it to learn to model the epidemic dynamics effectively.

\subsection{STGAT for epidemic prediction}
\begin{figure}[tb]\centering 
	\includegraphics[width=14cm]{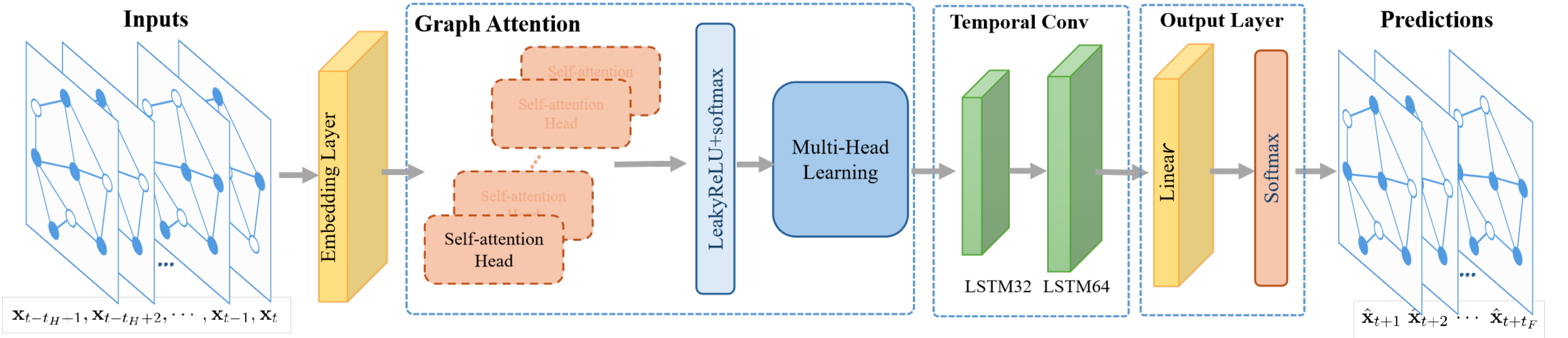}
	\caption{Architecture of spatio-temporal graph attention networks (STGAT).}
	\label{fig:stgat}
\end{figure}

We propose the Spatio-Temporal Graph Attention Network (STGAT) to address the problem of predicting epidemic dynamics by combining both spatial and temporal dependencies in a unified framework. The model is particularly suitable for scenarios where the spread of a disease is influenced by spatial interactions between individuals and the temporal progression of their historical states.
The proposed STGAT model for epidemic prediction integrates three key components: spatial modeling through graph attention mechanisms, temporal modeling via LSTM layers, and classification using fully connected layers, as illustrated in Figure \ref{fig:stgat}.

We first apply the embedding layer to map the discrete states of the nodes into a continuous dense vector space. Each node state $x_i(t)$ is mapped to a dense vector $e_i(t) \in \mathbb{R}^d$ through an embedding layer. Since the embedding layer processes the temporal state $x_i(t-t_H+1),...,x_i(t-1)$, the output leads to a matrix of $\hat{e}_i \in \mathbb{R}^{d\times t_H}$ where $t_H$ is the number of historical time steps.

\textbf{Graph Attention Network:}
The spatial component employs a Graph Attention Network (GAT) to model interactions between nodes in the network. 
In a traditional Graph Convolutional Network (GCN), the hidden state $h_i^{l+1}$ of a node $i$ at layer $l+1$ is updated by aggregating the features of its neighbors, usually using a simple average or sum. GAT further introduces an attention mechanism that assigns different weights (attention scores) to the neighbors based on their importance. By introducing an attention mechanism, GAT assigns weights to neighboring nodes based on their relevance, enabling the model to focus on the most influential connections. This dynamic weighting enhances the ability to capture spatial relationships, making the model adept at representing the complex topologies underlying epidemic spread.

Specifically, for each node $i$ and its neighbor $j$, an attention coefficient $\alpha_{ij}$ is computed based on their features, which indicates the importance of node $j$ in determining the updated feature of node $i$. The attention coefficient is computed as follows:
\begin{align}
	e_{ij} = \text{LeakyReLU}(a(\phi h_i^l, \phi h_j^l))
\end{align}
where $a(\cdot, \cdot)$ is the attention mechanism, $\phi$ is the learnable weight matrix for feature transformation.
Next, the softmax function is applied to normalize the attention scores across all neighbors of node $i$:
\begin{align}
	\alpha_{ij} = \frac{\exp(e_{ij})}{\sum_{k \in N(i)} \exp(e_{ik})}
\end{align}

Once the attention scores $\alpha_{ij}$ are computed, the node features are updated by aggregating the transformed features of the neighbors, weighted by their attention scores. To make the attention mechanism more robust, multi-head attention is employed. This means that the attention mechanism is applied multiple times in parallel (with different weight parameters), and the resulting outputs are either concatenated. The formula for multi-head attention is:
\begin{align}
	h_i^{l+1} = \bigg\|_{k=1}^K\sigma \left( \sum_{j \in Neigh(i)} \alpha_{ij} \phi h_j^l \right)
\end{align}
where $K$ is the number of attention heads, $||$ denotes concatenation of the outputs from each head, and $h_i^0 = \hat{e}_i(t)\in \mathbb{R}^d$ in the first layer of GAT.

The LSTM layers are then used to capture the temporal dependencies in the sequence of node states. In this model, after the GAT layer processes the spatial relationships, the output is passed through two LSTM layers. These layers learn the sequential patterns in the epidemic's spread by processing the temporal sequence of node states over multiple time steps, which follows
\begin{align}
	h_t = LSTM_2(LSTM_1(h^{l+1}))
\end{align}
where $h^{l+1}$ is the output of node feature after graph attention.  

Finally, the output layer is a fully connected layer with dropout probability that takes the final hidden state from the LSTM and maps it to a probability distribution over the possible classes, which follows the same formula (\ref{equ:softmax})(\ref{equ:argmax}) with the output layer in the above subsection. 

\section{Case study and discussion}
In this section, we present a series of comprehensive experiments designed to evaluate the efficacy of federated learning for epidemic prediction. We begin by detailing the simulation configurations and providing an assessment of the overall performance under centralized scenarios as a baseline for comparison. Subsequently, we explore the impact of varying the number of clients in the federated learning framework and introduce an efficacy energy indicator. Furthermore, we investigate the effect of the graph partitioning method, the epidemics, and the sensitivity of noise data.

\subsection{System configuration}
\subsubsection{Empirical spreading over the airports network}
The OpenFlights network\footnote{\url{https://openflights.org/}}, which documents regularly scheduled routes among airports worldwide, providing a realistic and well-structured representation of interactions between cities. In this study, we employ the airport network as a reasonable analogy to represent contact relationships. The behavior of epidemic spread over the flights network has been investigated by previous research \cite{findlater2018human}\cite{sokadjo2020influence}, highlighting their relevance in modeling global transmission patterns.
Moreover, this network structure mirrors real-world challenges in data interoperability, where administrative boundaries often hinder the direct exchange and utilization of epidemiological data between regions or countries. To construct a representative model, we extracted the top-ranking airports based on node degree, along with their corresponding airline connections, resulting in an undirected network comprising 600 nodes and 22,352 links. This approach ensures that the model captures the most significant transportation hubs, which are likely to play a critical role in facilitating epidemic transmission across regions.

We applied the Gillespie algorithm \cite{he2018spreading} to model the stochastic evolution of the epidemic on the network over time. The states of the nodes were recorded at fixed intervals, denoted as $\Delta t$, throughout the spreading process. 
One important consideration in this experiment is that, after a sufficiently long time, some certain epidemics may reach an extinction state where the majority of node states remain static, leading to a lack of meaningful variability for epidemic prediction.
Such scenarios are not meaningful for epidemic prediction, as they do not provide enough dynamic information for effective modeling. Therefore, we focus only the dynamic stage of the spreading process, where meaningful transmission interactions continue to occur, ensuring the dataset retains relevant information for prediction tasks.
The recorded node state sequences, as the dataset, was split into training, validation, and testing sets, corresponding to 20\%, 10\%, and 70\% of the data from the beginning of the recorded time series, respectively. 
To rigorously evaluate the model’s ability to predict critical transitions, the training and validation datasets were limited to the early stages of the spread, with minimal data included near the inflection point or peak \cite{he2018spreading}. This setting allows for an assessment of the model's effectiveness in forecasting pivotal turning points in epidemic dynamics.
Additionally, the training data were randomly shuffled during the training process to promote the learning of generalized patterns from spreading dynamics.

\subsubsection{Experiment setting}
All experiments were conducted on a server equipped with 4 NVIDIA GeForce RTX4090 GPUs. The past time window is 10 time steps (i.e., $t_H=10$ observed data points) and they are used to forecast traffic speed in the next $t_F=10$ time steps.
The LSTM model with 64 hidden units was trained using the Adam optimizer for 1000 epochs. where the initial learning rate is $2e^{-4}$ with a weight decay of $5e^{-5}$.
The STGAT model was also trained using the Adam optimizer with an initial learning rate of $2e^{-4}$, a weight decay of $5e^{-5}$, and a batch size of 32.  A single graph attention layer with 8 heads was employed to implement the multi-head attention mechanism. The temporal component of the STGAT model consisted of two LSTM layers with 32 and 64 hidden units, respectively. To stabilize the learning process, batch normalization and Xavier parameter initialization were used. Additionally, dropout and an early stopping strategy were implemented to mitigate overfitting.

To evaluate the performance of the proposed models for node state prediction, we employ several metrics, including cross-entropy (CE), accuracy (Acc), and F1 score.
In addition to node-level predictions, we also analyze the dynamic behavior of the epidemic by focusing on the prevalence $y_I(t)$, which represents the fraction of infected nodes at time and and reflects the overall trend of infection spread. 
Let $\hat{y}_I(t)$ denote the predicted prevalence at time $t$, we employ Root Mean Square Error (RMSE), Mean Absolute Error (MAE) as complementary metrics to provides an intuitive measure of the model’s performance to trace the overall behavior, which are defined as follows
\begin{align}
	\text{RMSE} &= \sqrt{\frac{1}{T} \sum_{i=1}^{T} \big( y_I(t) - \hat{y}_I(t) \big)^2} \\
	\text{MAE} &= \frac{1}{T} \sum_{i=1}^{T} \left| y_I(t) - \hat{y}_I(t) \right|
\end{align}
where $T$ is the total number of time steps in the evaluation period.

\subsubsection{Overall performance in centralized scenario}
In this study, we first evaluate the performance of our models for various epidemic types with predetermined parameters under a centralized learning scenario, where the models are trained using the complete state information of all nodes, as presented in Table \ref{tlb:central}. The results reveal that no single prediction model consistently outperforms the other across different epidemiological models, while the comparison results of different metrics for model performance are consistent. For instance, the LSTM model demonstrates better performance in predicting SIS, SIR, SEIR, SIRVS epidemics, whereas STGAT excels in the nmSIS, SIStv, SIRS processes. 

\begin{table}[h!]
	\centering
	\footnotesize
	\renewcommand{\arraystretch}{1.4}
	\scalebox{0.9}{
	  \begin{tabular}{@{\extracolsep{\fill}}c|ccccc|ccccc}
			\hline
			\hline
			\multicolumn{1}{c|}{\textbf{Model}} & \multicolumn{5}{c|}{\textbf{LSTM}} & \multicolumn{5}{c}{\textbf{STGAT}} \\
			\hline
			\textbf{Metrics} & \textbf{CE} & \textbf{Accuracy} & \textbf{F1 score} & \textbf{RMSE} & \textbf{MAE} & \textbf{CE} & \textbf{Accuracy} & \textbf{F1 score} & \textbf{RMSE} & \textbf{MAE} \\
			\hline
			SIS & \textbf{0.0157} & \textbf{0.9982} & \textbf{0.9978} & \textbf{0.3689} & \textbf{0.0321} & 0.0182 & 0.9954 & 0.9605 & 0.4908 & 0.0450 \\
			
			nmSIS & 0.1503 & 0.9769 & 0.9769 & 1.5843 & 0.4476 & \textbf{0.1429} & \textbf{0.9801} & \textbf{0.9791} & \textbf{1.2305} & \textbf{0.2828} \\
			
			SIStv & 0.0679 & 0.9897 & 0.9896 & 0.7967 & 0.1003 & \textbf{0.0542} & \textbf{0.9979} & \textbf{0.9974} & \textbf{0.7354} & \textbf{0.0941} \\
			
			SIRS & 0.0249 & 0.9963 & 0.9879 & 0.3882 & 0.0231 & \textbf{0.0212} & \textbf{0.9963} & \textbf{0.9887} & \textbf{0.3674} & \textbf{0.0214} \\
			
			SIR & \textbf{0.0392} & \textbf{0.9987} & \textbf{0.9978} & \textbf{0.2358} & \textbf{0.0143} & 0.0540 & 0.9983 & 0.9967 & 0.2606 & 0.0163 \\
			
			SEIR & \textbf{0.0350} & \textbf{0.9921} & \textbf{0.9920} & \textbf{0.4341} & \textbf{0.0489} & 0.0399 & 0.9904 & 0.9899 & 0.6027 & 0.0524 \\
			
			SIRVS & \textbf{0.0671} &\textbf{ 0.9904} & \textbf{0.9876} & \textbf{0.4934} & \textbf{0.0356} & 0.0895 & 0.9867 & 0.9843 & 0.5638 & 0.0419 \\
			\hline
			\hline
		\end{tabular}
	}
	\caption{Performance comparison of two models, i.e., LSTM and STGAT, for different epidemiological models in the centralized learning scenarios. The metrics CE, Accuracy, F1, RMSE, MAE are applied, where the bold denotes the better results for each epidemic model.}
	\label{tlb:central}
\end{table}

\begin{figure}
    \centering
    \includegraphics[width=0.65\linewidth]{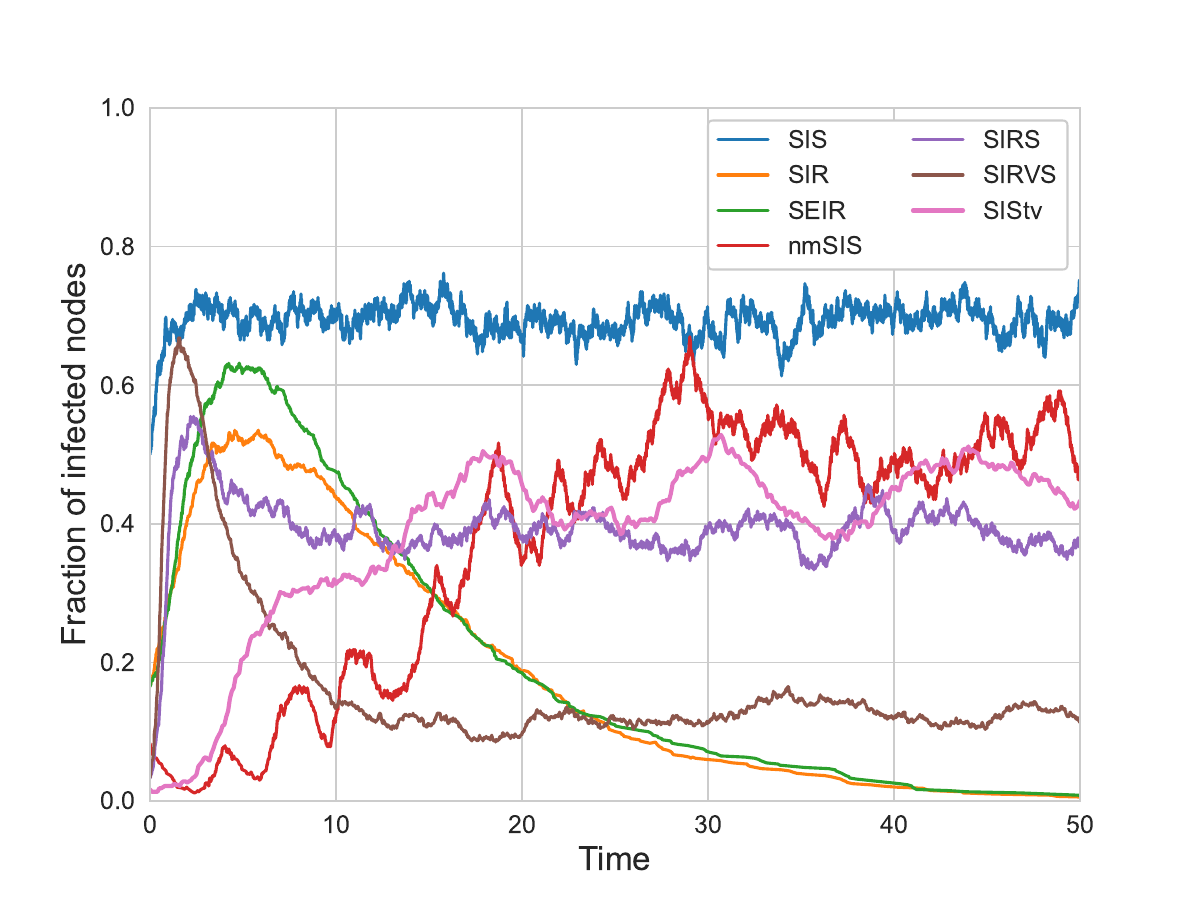}
    \caption{Illustrations of the fraction of infected nodes as a function of time for various epidemic processes, including SIS, SIR, SEIR, SIRVS, nmSIS, SIStv, and SIRS.}
    \label{fig:prevalence}
\end{figure}

This variation in performance can be partially explained by the prevalence patterns of these epidemics, as illustrated in Figure \ref{fig:prevalence}. Epidemics such as SIS, SIR, SEIR, and SIRVS usually display a transition period and tend to reach a stable state, whereas others such as nmSIS, SIStv, and SIRS exhibit periodic fluctuations over time after transition period. Given that our training and validation datasets primarily utilize the early stages of epidemic spread, with limited data before the inflection point or peak, the LSTM model excels at learning the general trend of the epidemic spread with less data. However, it may struggle to effectively capture periodic patterns due to its reliance on sequential temporal information alone.

In contrast, the STGAT model leverages topological information, making it more adept at identifying pivotal points and adapting to periodic fluctuations. The incorporation of graph-based attention mechanisms enables the STGAT model to exploit structural dependencies among nodes, providing a precise tracking of transmission dynamics. This ability provides advantages in scenarios involving cyclical or intricate epidemic patterns, where capturing both spatial and temporal dependencies is critical for accurate prediction. 

Epidemic prediction in this paper is framed as a classification task, where nodes are categorized into one of several states. However, within a short time interval corresponding to the observed time step, the majority of nodes remain unchanged, with only a small subset undergoing dynamic transitions. For instance, in an SIR process, nodes in the recovered (R) state are inherently static and will persist in the R state indefinitely, making their classification trivial.
As a result, accuracy, which quantifies the proportion of correctly classified nodes, tends to be disproportionately influenced by the majority of static nodes. This may lead to inflated accuracy values and negligible numerical differences between scenarios, failing to provide meaningful insights into the model's performance on dynamically changing nodes.
To address these limitations, we complement the reciprocal of cross-entropy (1/CE) and the RMSE as the evaluation metrics in the subsequent parts. These metrics offer a more nuanced assessment by emphasizing both the predictive uncertainty and the magnitude of errors, ensuring a fair and robust comparison across different scenarios.

\subsection{Efficacy of federated epidemic prediction}
This subsection evaluates the efficacy of the proposed approach for federated epidemic prediction across a range of different scenarios. We examine the relationship between model performance and the number of clients, and then propose a general metric, i.e, efficacy energy. Based on the experiment setting mentioned earlier, all federated frameworks are trained for 200 epochs of client-server communication rounds with 5 epochs of local training rounds, and the proximal term $\mu$ for FedProx is set to 0.01 as default. 

\subsubsection{Performance versus number of clients}
In the federated setting, the entire network comprising $N$ nodes is partitioned evenly among $M$ clients based on the indices of the nodes (i.e. airport index), where each client is responsible for a subnetwork containing approximately $[N/M]$ nodes.
Let $\alpha_m$ be the performance metric of the client $m$, we define the average metric $\bar{\alpha}[M] = \frac{1}{M}\sum_{i=1}^M \alpha_m$ among $M$ clients to quantifies the effectiveness performance in a specific federated scenario. 
The average metric $\bar{\alpha}[M]$ is inherently influenced by three primary factors: the choice of the model $\mathcal{M}$ (e.g., LSTM or STGAT), the aggregation method $\mathcal{F}$ applied in the federated framework (e.g., FedAvg or FedProx), and the underlying epidemic process $\mathcal{P}$ (e.g., SIS, SIR, etc.). These factors collectively impact the prediction accuracy and robustness of the federated learning approach across diverse scenarios.

\begin{figure}[tb]\centering 
	\subfloat[LSTM for nmSIS under 4 clients by $\alpha=Acc$]
	{\includegraphics[width=8cm]{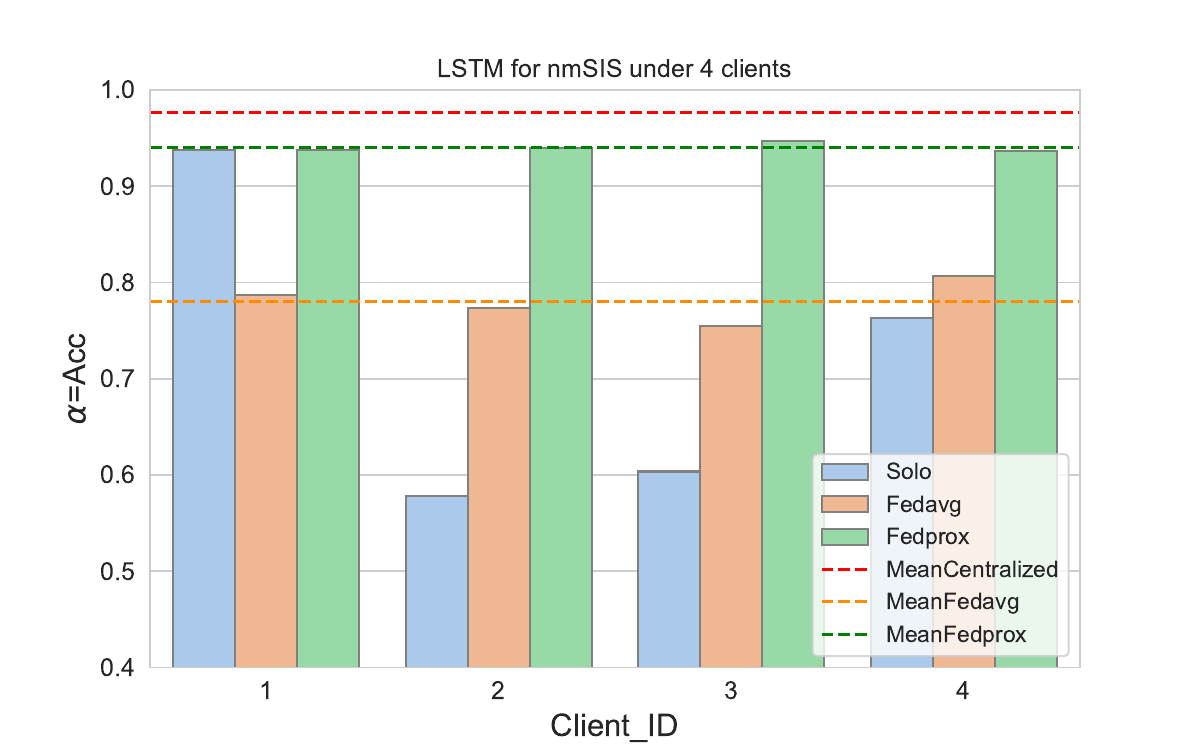}} \hfil
	\subfloat[STGAT for nmSIS under 4 clients by $\alpha=Acc$]
	{\includegraphics[width=8cm]{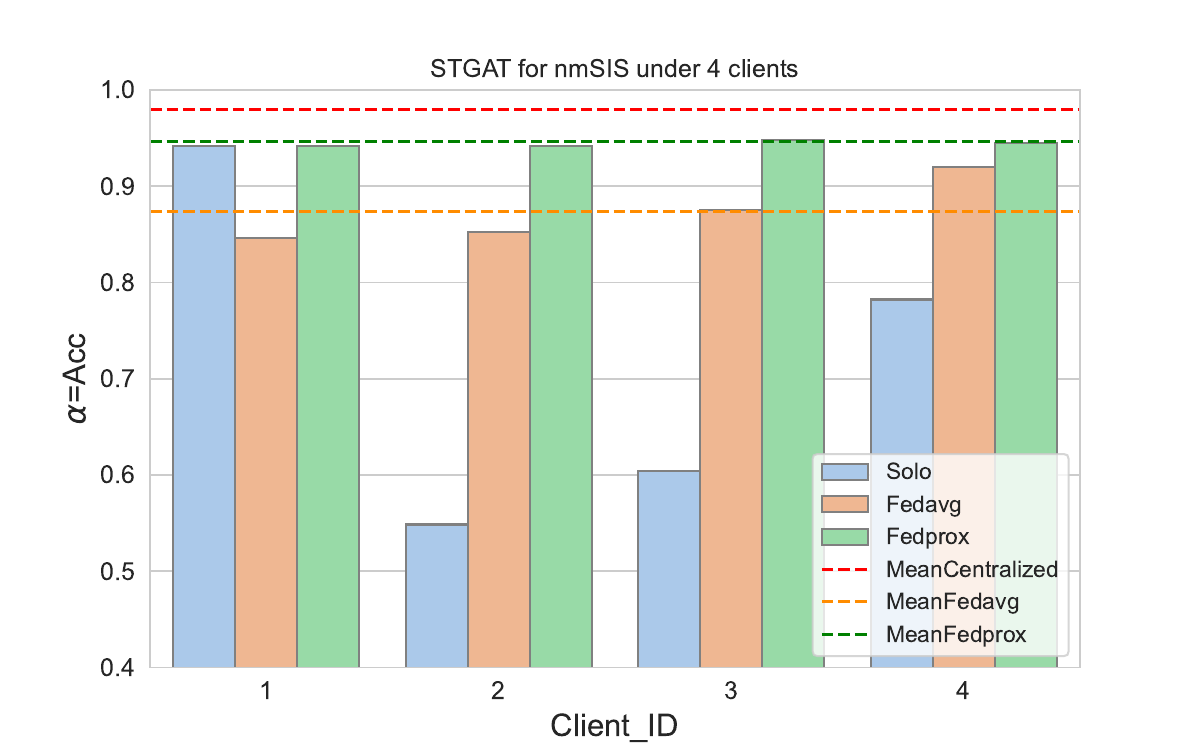}}\\
	\subfloat[LSTM for nmSIS under 8 clients by $\alpha=Acc$]
	{\includegraphics[width=8.5cm]{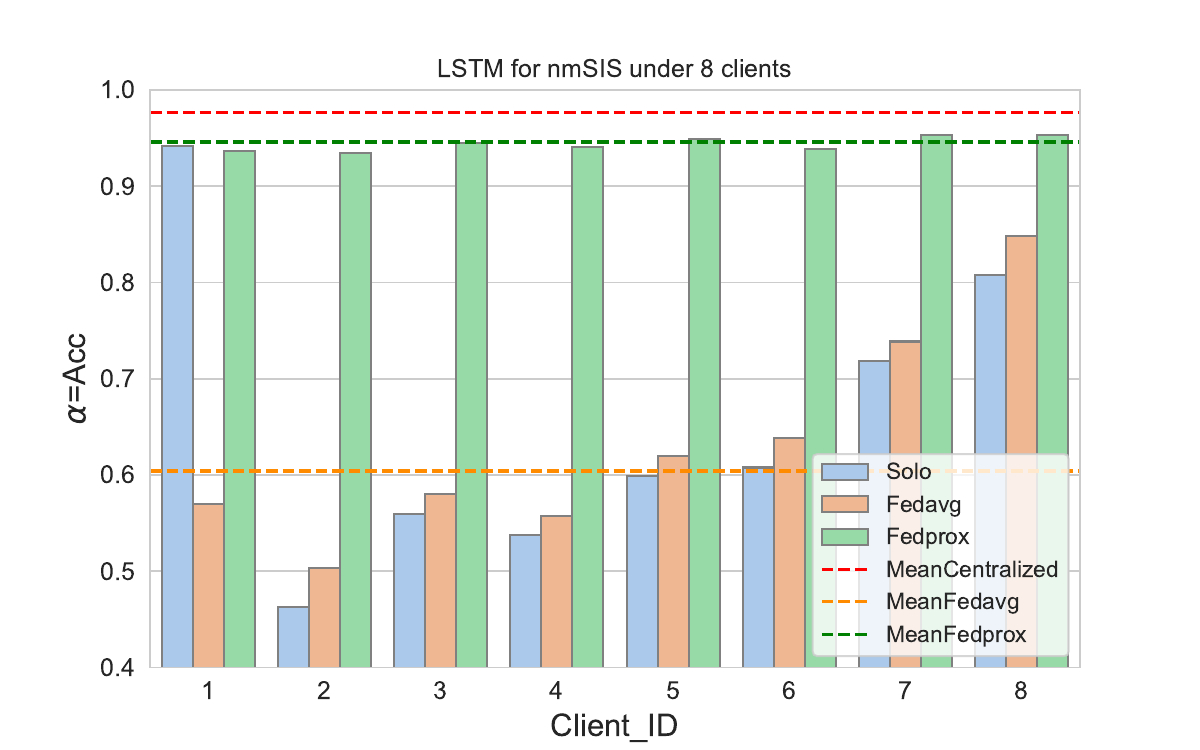}} \hfil
	\subfloat[STGAT for nmSIS under 8 clients by $\alpha=Acc$]
	{\includegraphics[width=8.5cm]{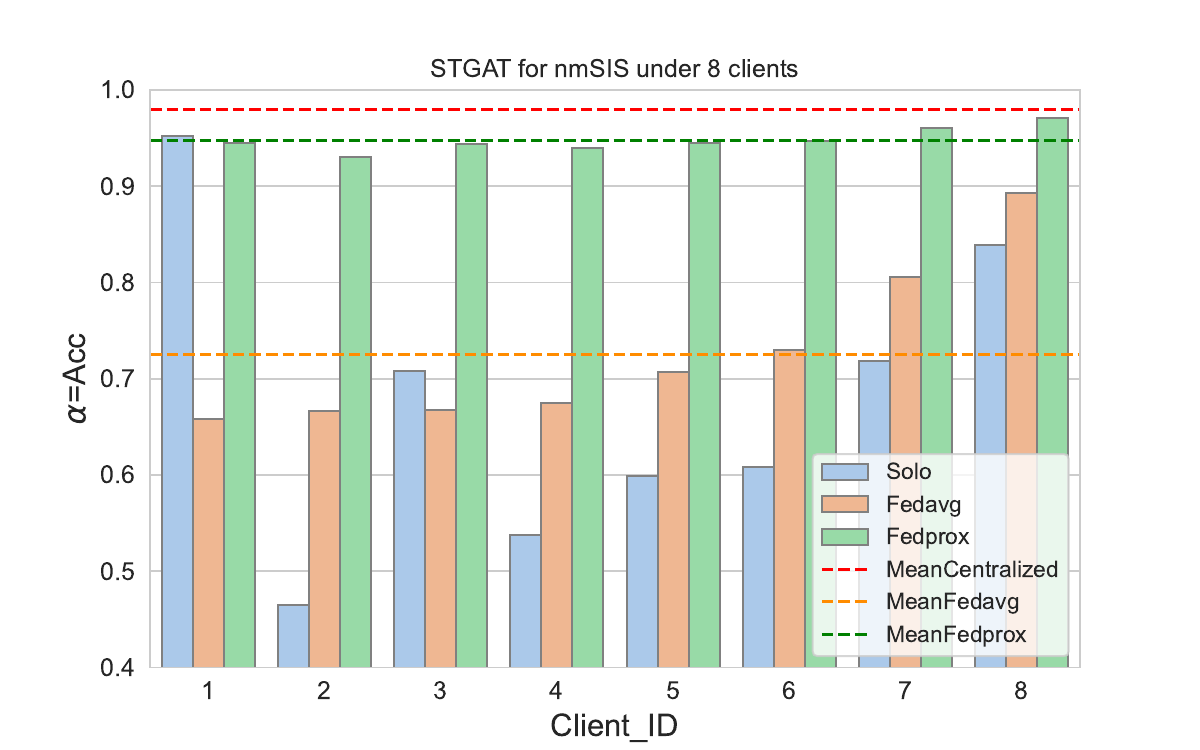}}
	\caption{Comparison of centralized learning and federated learning for an exampled nmSIS process under 4 or 8 clients by the accuracy metric $\alpha=Acc$. The bar represents the performance under solo learning, Fedavg learning and Fedprox learning. The horizontal lines indicate the mean metric under centralized learning, Fedavg learning and Fedprox learning. }
	\label{fig:solo_compare}
\end{figure}

Figure \ref{fig:solo_compare} illustrates the performance with the accuracy metric $\alpha=Acc$ of each clients in the FL system for an exampled nmSIS process.
Federated learning demonstrates significant improvements over solo learning, where each client independently trains on its local dataset. For both LSTM and STGAT models, most clients benefit from the collaborative knowledge sharing enabled by federated learning, resulting in enhanced prediction accuracy and improved generalization.
However, a small subset of clients exhibits marginally lower performance, potentially due to data heterogeneity or aggregation effects in the federated setting. Among the aggregation methods, FedProx consistently outperforms FedAvg, achieving performance closer to centralized learning. The superior performance highlights FedProx's advantage to mitigate data heterogeneity and stabilize the training process, establishing it as a robust choice for federated epidemic prediction.

\subsubsection{The efficacy energy for various epidemics}
\begin{figure}[tb]\centering 
	\subfloat[$\alpha=1/CE$ for nmSIS]
	{\includegraphics[width=8.5cm]{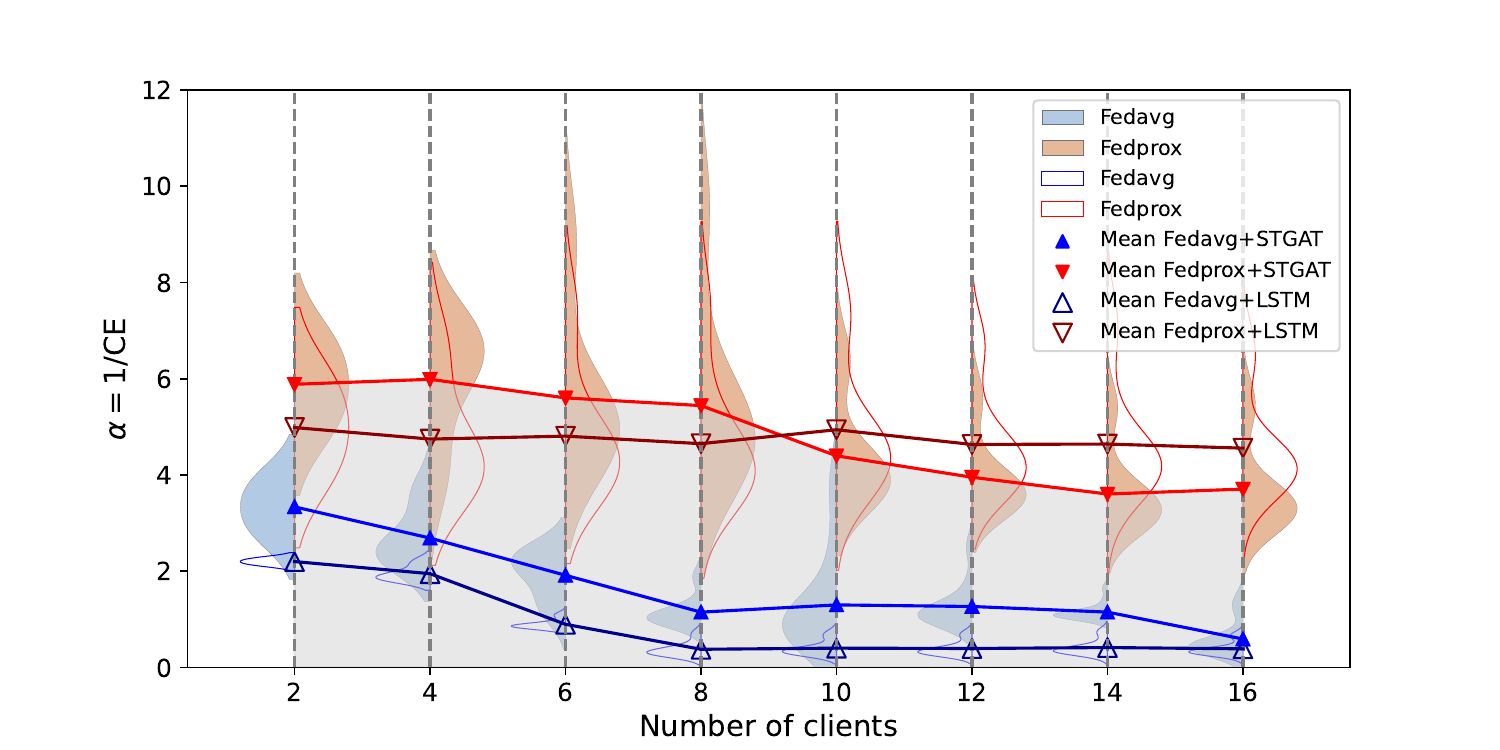}} \hfil
	\subfloat[$\alpha=$RMSE for nmSIS]
	{\includegraphics[width=8.5cm]{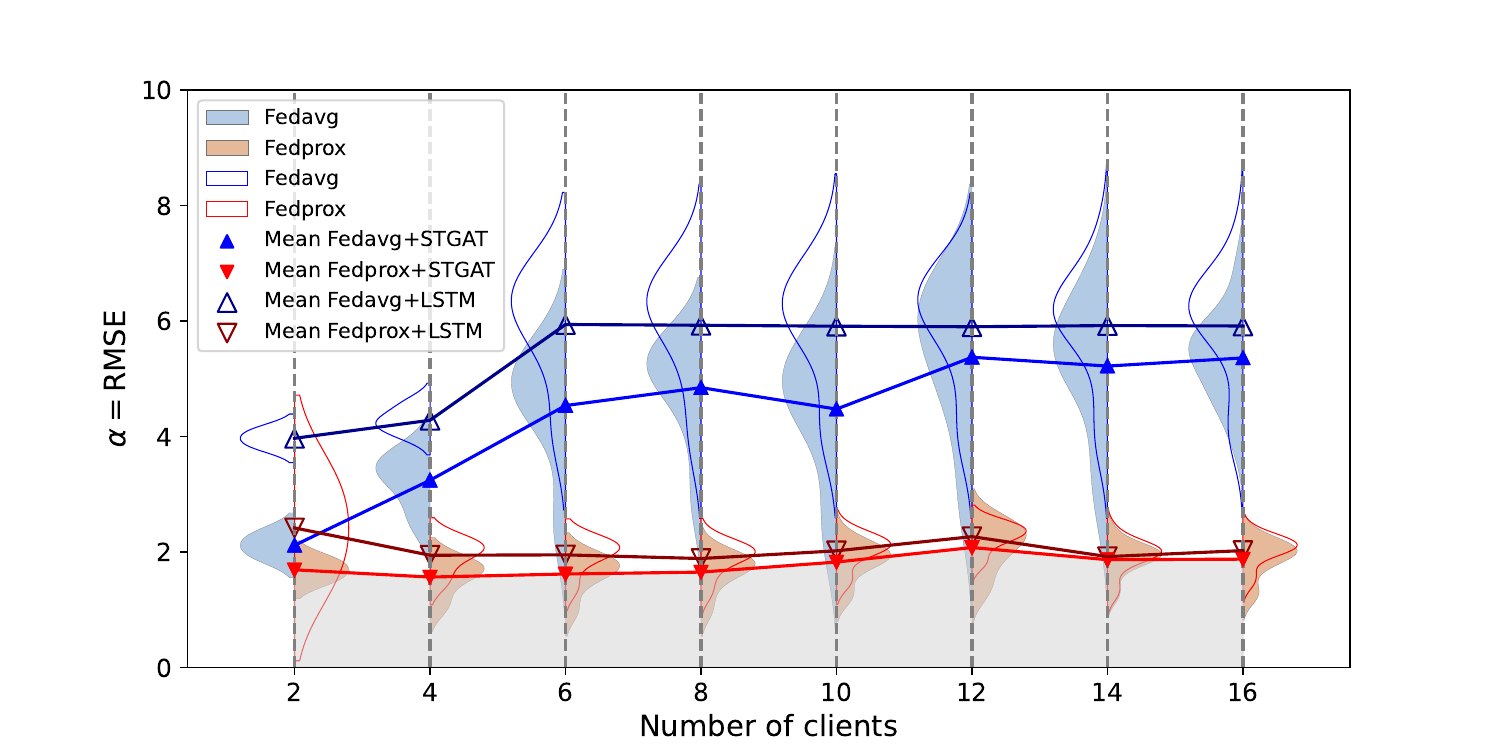}}
	\caption{Violin plots of the metrics, e.g. $1/CE$ and RMSE, as a function of the number of clients $M$ in a nmSIS federated learning. The histograms represent the metrics among all clients under different scenarios, and the line represent the mean of the metric $\bar{\alpha}$. The gray surface illustrates the efficacy energy of the scenario of STGAT and Fedprox for nmSIS with $M_0=16$.}
	\label{fig:violin_nmsis}
\end{figure}

Unlike previous studies conducted with a fixed number of clients, which offer limited insights, this work investigates the impact of varying client numbers on the efficacy of federated epidemic prediction. Figure \ref{fig:violin_nmsis} presents representative performance metrics, such as the reciprocal of cross-entropy ($1/CE$) and the RMSE, as functions of the number of clients within the federated learning framework.
The average performance metric $\bar{\alpha} = E[1/CE]$ and $\bar{\alpha} = E[RMSE]$, exhibits an overall degradation trend as the number of clients increases, probably due to the reduced training sample size available per client as the increasing number of clients. Beyond a certain threshold of clients ($M$), the metric stabilizes at a relatively poor level, reflecting the inherent limitations in model learnability caused by reduced local datasets.
Among the aggregation methods evaluated, FedProx outperforms FedAvg, particularly as the number of clients grows. While both methods experience performance degradation, FedProx exhibits a slower decline, highlighting its ability to mitigate the challenges of data heterogeneity and limited client datasets.

We observe that no single model consistently outperforms the other across all client configurations, as illustrated by the red lines by FedProx in Figure \ref{fig:violin_nmsis}(a). In our case, the lower dependency on data volumes of the LSTM might leads to a less performance degradation for a larger number of clients ($M>9$). The optimal choice of learning scenarios depending on the number of clients and the underlying data distribution.
In practical federated learning scenarios, the participating clients is often difficult to predetermine or may vary dynamically over time. Thus, assessments under fixed client configurations fail to provide a comprehensive evaluation of the adaptability of FL systems in real-world applications.
To address this variability and uncertainty, we define the efficacy energy $\eta$ as the mean of the cumulative average performance metric $\bar{\alpha}$ across an increasing number of clients, which follows:
\begin{align}
	\eta(\alpha, \mathcal{M},\mathcal{F},\mathcal{G},\mathcal{P}) = \frac{1}{M_0-2}\sum_{K=2}^{M_0} \bar{\alpha}[M]
\end{align}
where $M_0$ is an estimated upper-limit number of clients. 
By aggregating performance across diverse client configurations, the efficacy energy $\eta$ serves as a robust indicator of the system's effective resilience under uncertain conditions. which allows for a more comprehensive evaluation of the robustness on the efficacy of federated learning in real-world applications.

\begin{figure}
	\centering
	\includegraphics[width=0.9\linewidth]{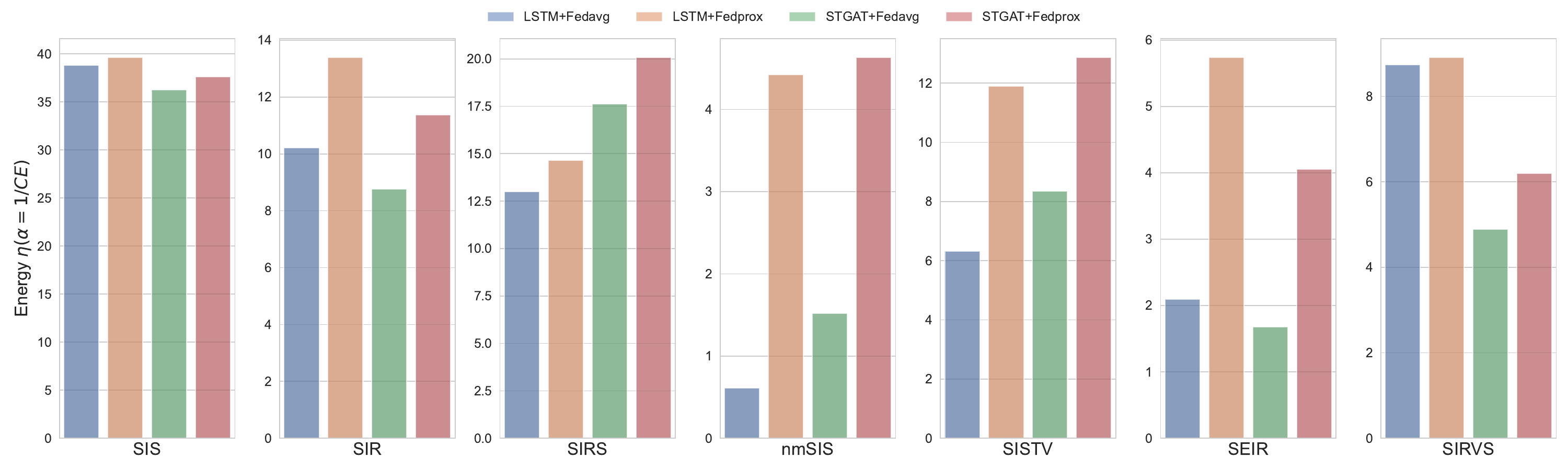}
	\caption{Performance energy $\eta(\alpha=1/CE)$ for 7 epidemic process under different prediction models and aggregation approaches, with the upper-limit number of clients $M_0=16$.}
	\label{fig:7dataset_1ce}
\end{figure}

Figure \ref{fig:7dataset_1ce} shows the efficacy energy $\eta(\alpha=1/CE)$ for 7 epidemic process under different prediction models (LSTM and STGAT) and aggregation methods (FedAvg and FedProx). 
Overall, FedProx consistently achieves higher energy values compared to FedAvg for both LSTM and STGAT models. This observation underscores FedProx's ability to address the challenges posed by increasing client heterogeneity, resulting in more stable and reliable performance across diverse scenarios.
For epidemic processes characterized by dynamic or fluctuating behaviors, such as nmSIS and SIStv, the combination of STGAT and FedProx exhibits superior efficacy energy. This result highlights the advantage of leveraging graph-based topological information to effectively capture the complex spatio-temporal patterns inherent in such processes.
Conversely, for simpler epidemic dynamics like SIR, the combination of the LSTM and FedProx performs comparably better than the STGAT, which suggests that temporal models alone may suffice for relatively straightforward transmission dynamics, without the effort of complex graph-based spatial information.

\subsection{Performance under different graph partitioning}
\begin{figure}[tb]\centering 
	\subfloat[LSTM for SIS]
	{\includegraphics[width=5cm]{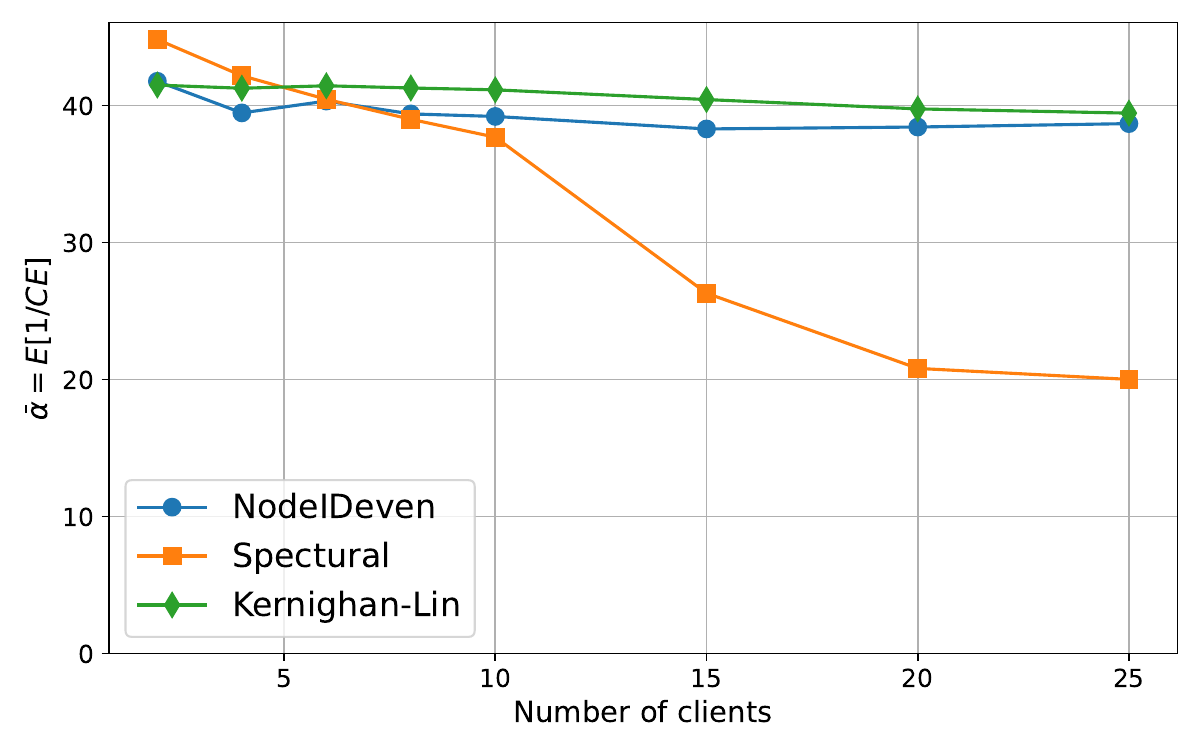}} \hfil
	\subfloat[LSTM for nmSIS]
	{\includegraphics[width=5cm]{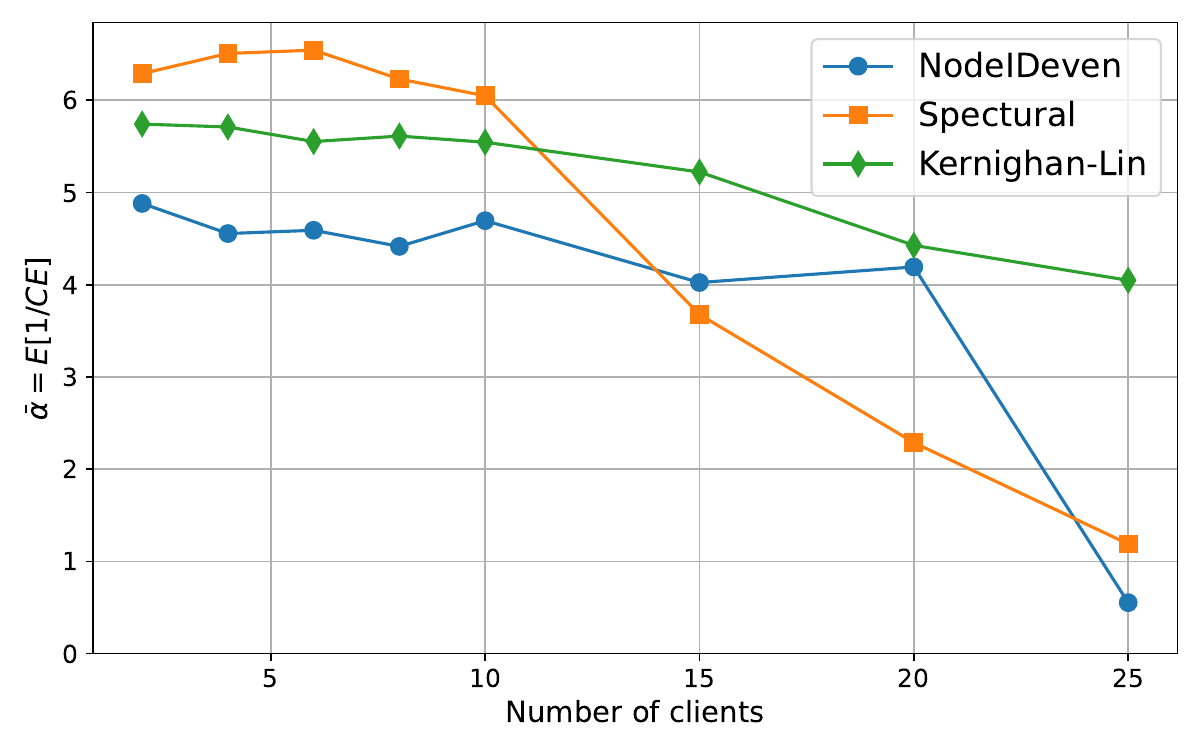}} \hfil
	\subfloat[LSTM for SEIR]
	{\includegraphics[width=5cm]{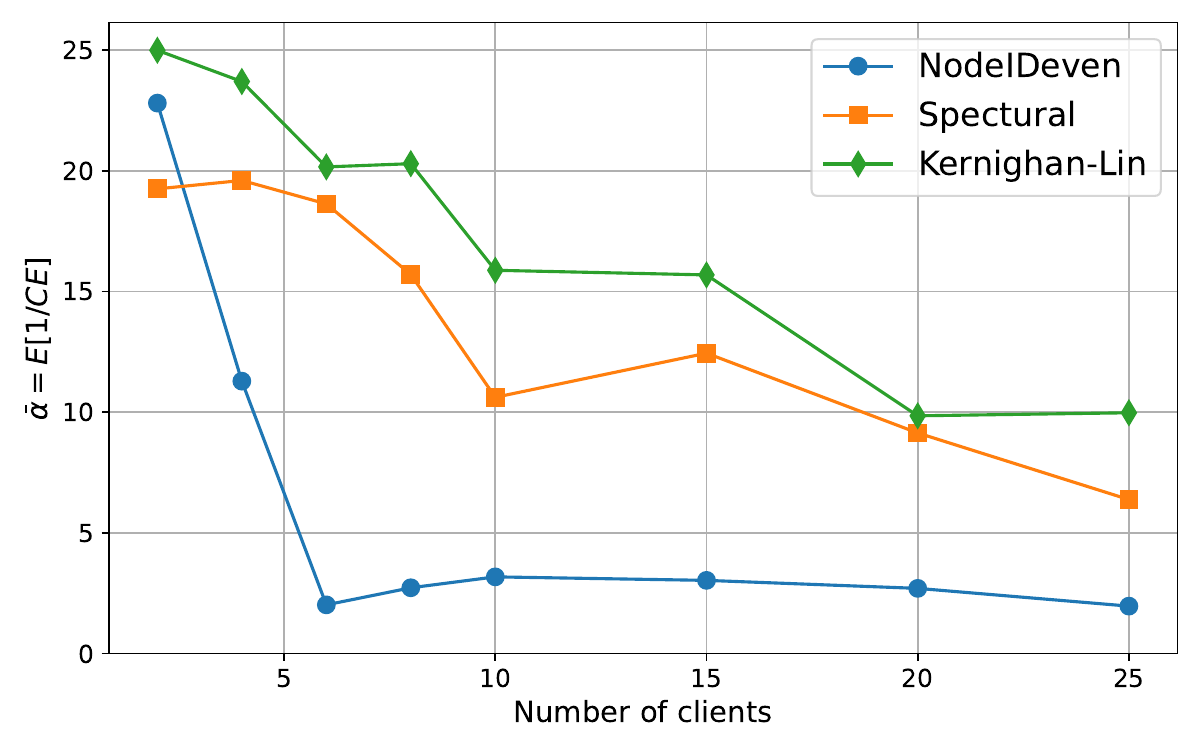}}\\
	\subfloat[STGAT for SIS]
	{\includegraphics[width=5cm]{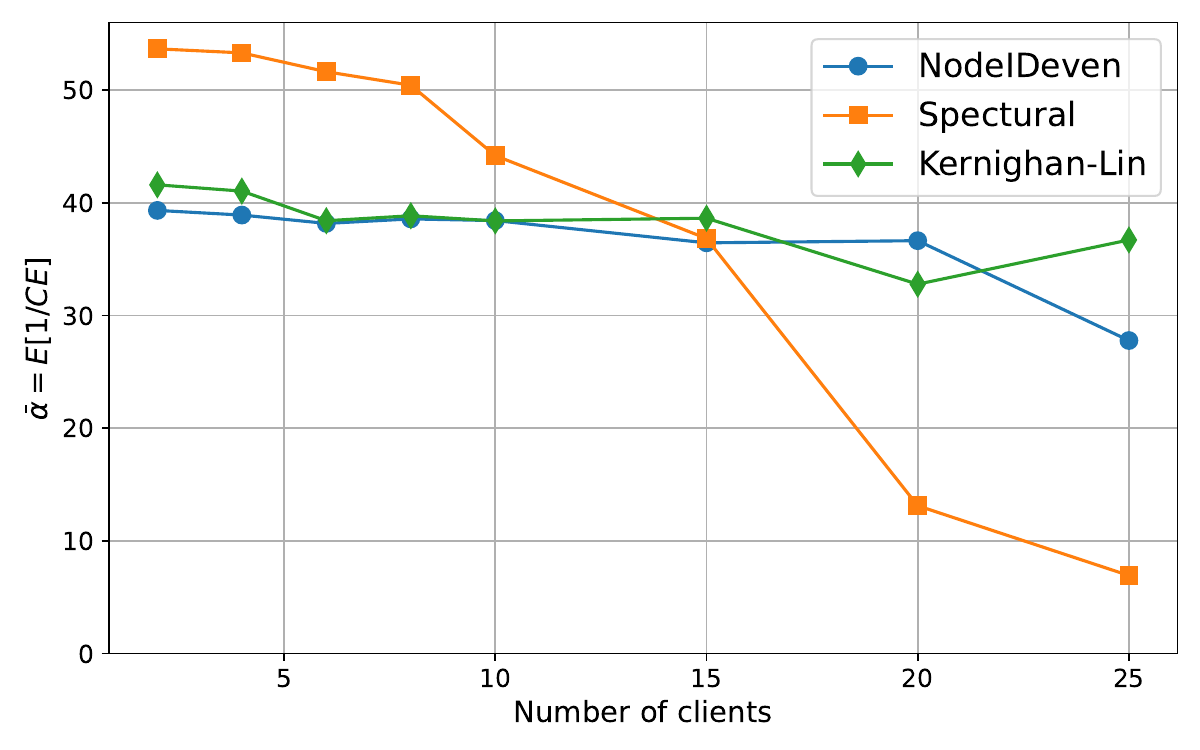}} \hfil
	\subfloat[STGAT for nmSIS]
	{\includegraphics[width=5cm]{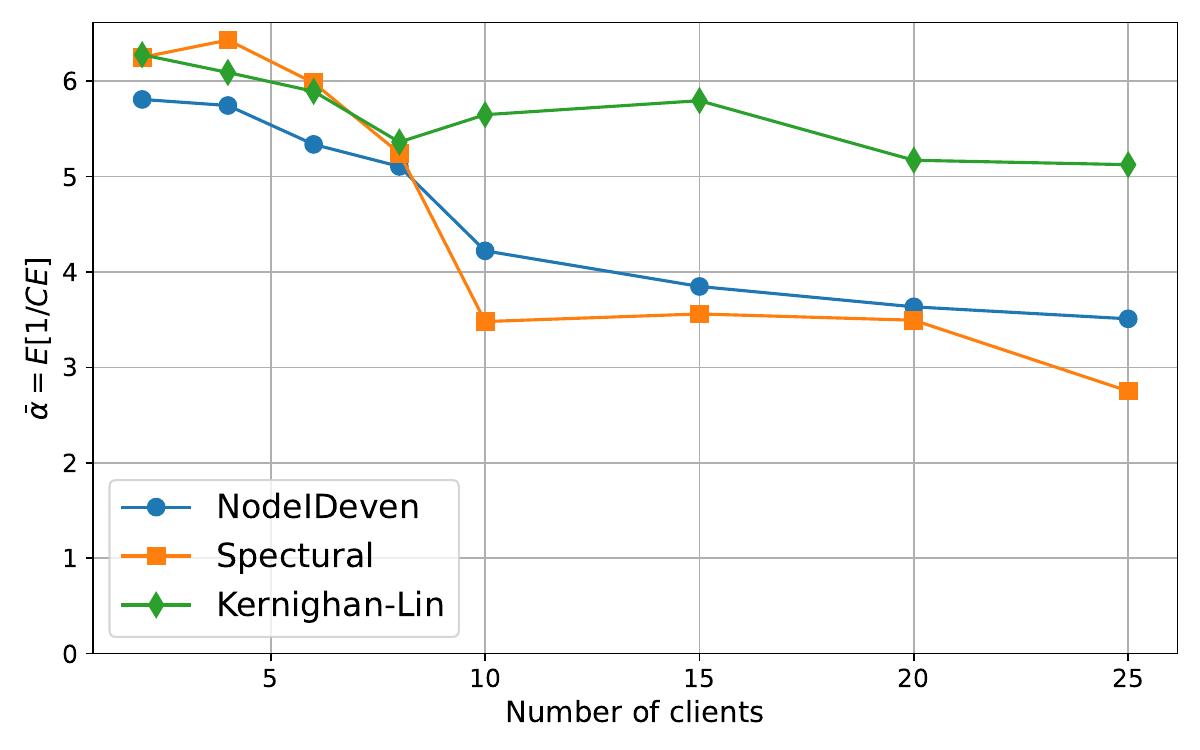}} \hfil
	\subfloat[STGAT for SEIR]
	{\includegraphics[width=5cm]{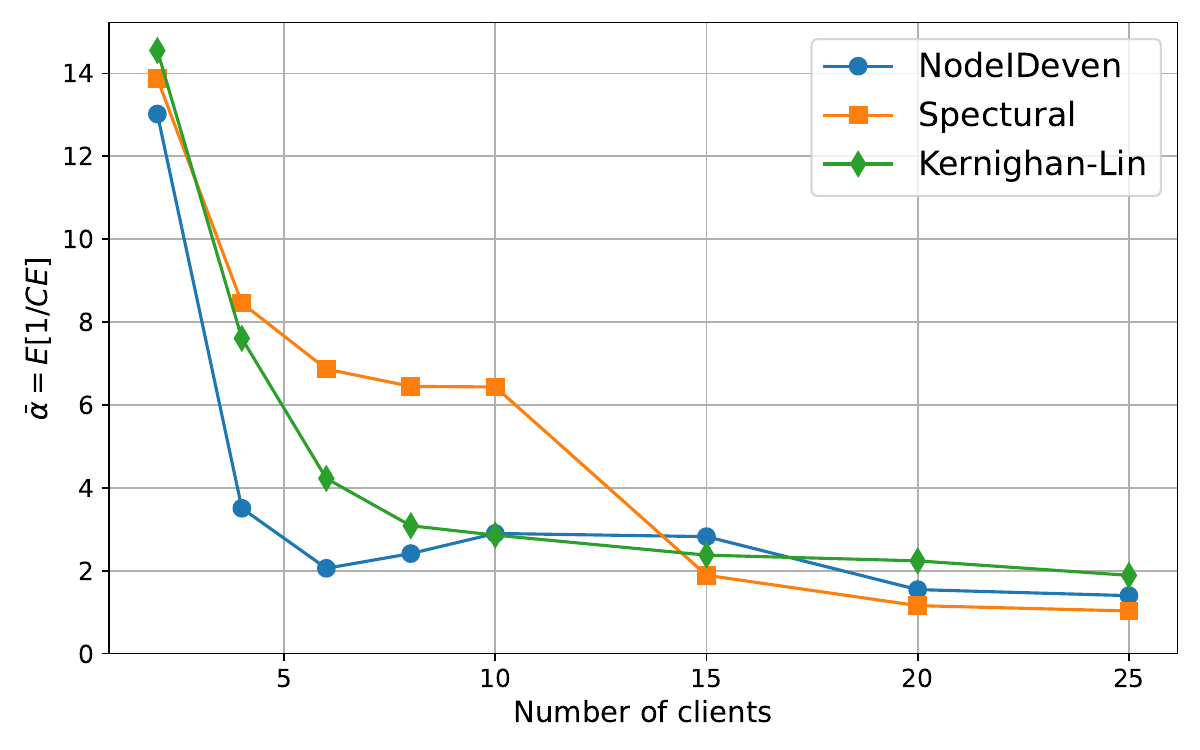}}
	\caption{Performance metric $\bar{\alpha}=E[1/CE]$ as a function of the number of clients $M$ under different graph partition $\mathcal{G}$ for three epidemic cases, e.g., SIS, nmSIS and SEIR. }    
	\label{fig:graph_partition}
\end{figure}

\begin{figure}
	\centering
	\includegraphics[width=0.8\linewidth]{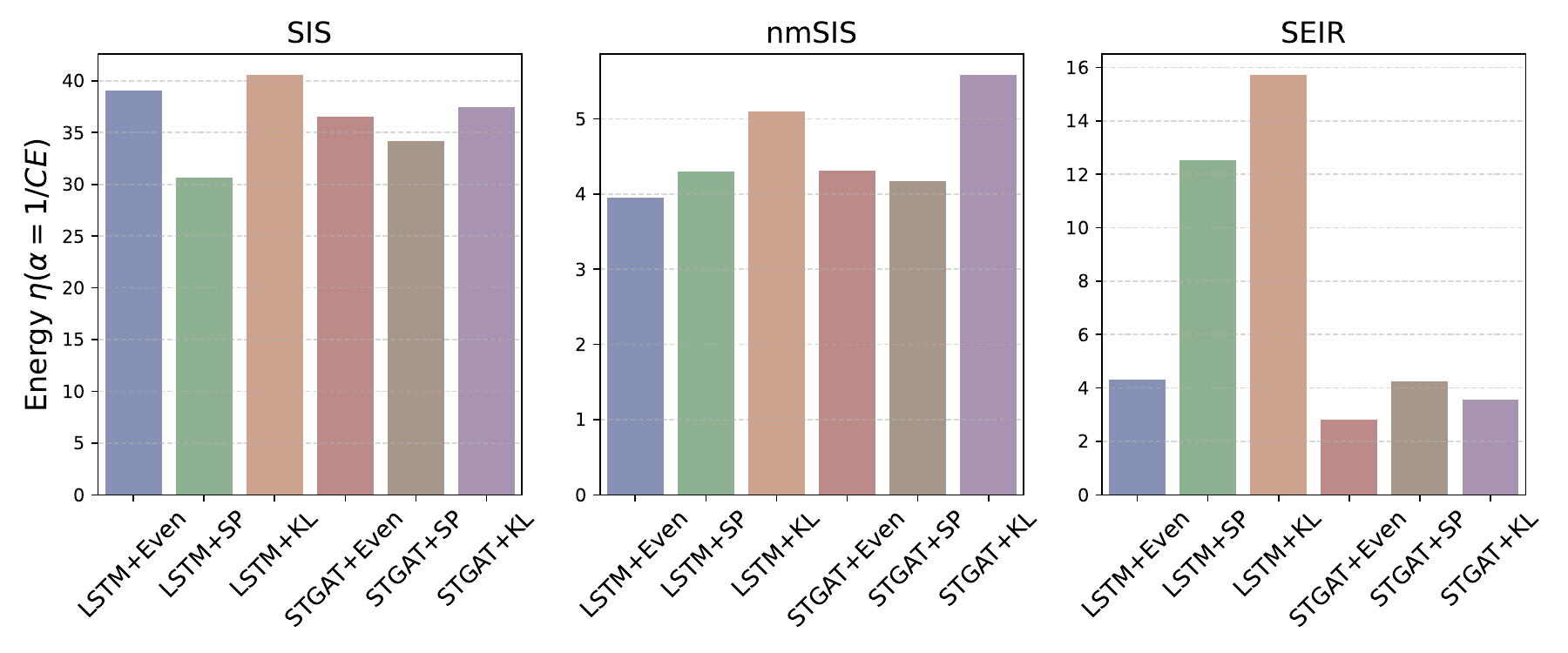}
	\caption{Efficacy energy $\eta$ as the metric $\alpha=1/CE$ of the reciprocal CE for different federated scenarios. The combination of prediction model (the LSTM and the STGAT) and three graph partitioning methods for three typical epidemics are investigated. The upper-limit number of client is $M_0=25$.}
	\label{fig:cluster_energy}
\end{figure}

We further investigate the impact of different graph partitioning methods, denoted as $\mathcal{G}$, on the performance of federated learning for epidemic prediction. In addition to the even partitioning by node index, we introduce two other graph partitioning methods: spectral clustering and Kernighan-Lin partitioning.
Spectral clustering\cite{fiedler1973algebraic} is a graph partitioning method that leverages the eigenvectors of the graph Laplacian matrix to group nodes into clusters. This approach captures the underlying structure of the graph, ensuring that nodes with strong connectivity are assigned to the same subnetwork. On the other hand, Kernighan-Lin (KL) partitioning\cite{kernighan1970efficient} is a heuristic-based algorithm that minimizes edge-cut cost between clusters while maintaining balance in the size of the partitions.

Figure \ref{fig:graph_partition} illustrates the performance metric $\bar{\alpha}=E[1/CE]$ as a function of the number of clients $M$ for three typical epidemic cases, e.g., SIS, nmSIS and SEIR. We observe that spectral clustering performs best for a small number of clients, likely because epidemic spreading often exhibits localization phenomena \cite{goltsev2012localization}. In such cases, nodes within a localized cluster tend to exhibit feature consistency, making the partition more learnable for individual clients. 
However, as the number of clients increases, the performance of spectral clustering degrades sharply, while the other methods, such as even partitioning by node index and KL partitioning, demonstrate superior performance. This degradation is likely due to the uneven data distribution among clients and extremely sparse dateset within some clients by spectral clustering, which reduce learnability. For a larger number of clients, the uniformity of data volume becomes more critical for effective learning.

Noticeably, Kernighan-Lin partitioning demonstrates robust and consistently good performance with a large performance energy $\eta$, as depicted in Figure \ref{fig:cluster_energy}. By balancing the uniformity of data volume across clients and the consistency of data features within partitions, KL partitioning achieves favorable results across diverse client configurations.
The above findings highlight the importance of selecting appropriate data segmentation methods for federated epidemic prediction, depending on both the number of clients and the characteristics of the epidemic spreading process.

\subsection{Performance under different phases of epidemics}
Epidemics on networks are known to exhibit phase transition phenomena, characterized by extinction states, multiple metastable states, or bifurcation behaviors. 
To explore the impact of different phases on federated prediction performance, we consider two typical epidemic models, SIS and SIRS, whose effective infection rates are defined as $\tau=\beta/\delta$ and $\tau=\beta/(\delta+\omega)$, respectively.
The epidemic threshold is estimated using the first-order mean-field approximation, given by $\tau_c^{(1)}= 1/ \lambda_1(A)$, where $\lambda_1(A)$ is the largest eigenvalue of the adjacency matrix $A$.
When the effective infection rate $\tau>\tau_c$, the spreading process tends toward a metastable state, while for $\tau<\tau_c$, the epidemic transitions to an extinction state.

\begin{figure}[tb]\centering 
	\subfloat[SIRS]
	{\includegraphics[width=8cm]{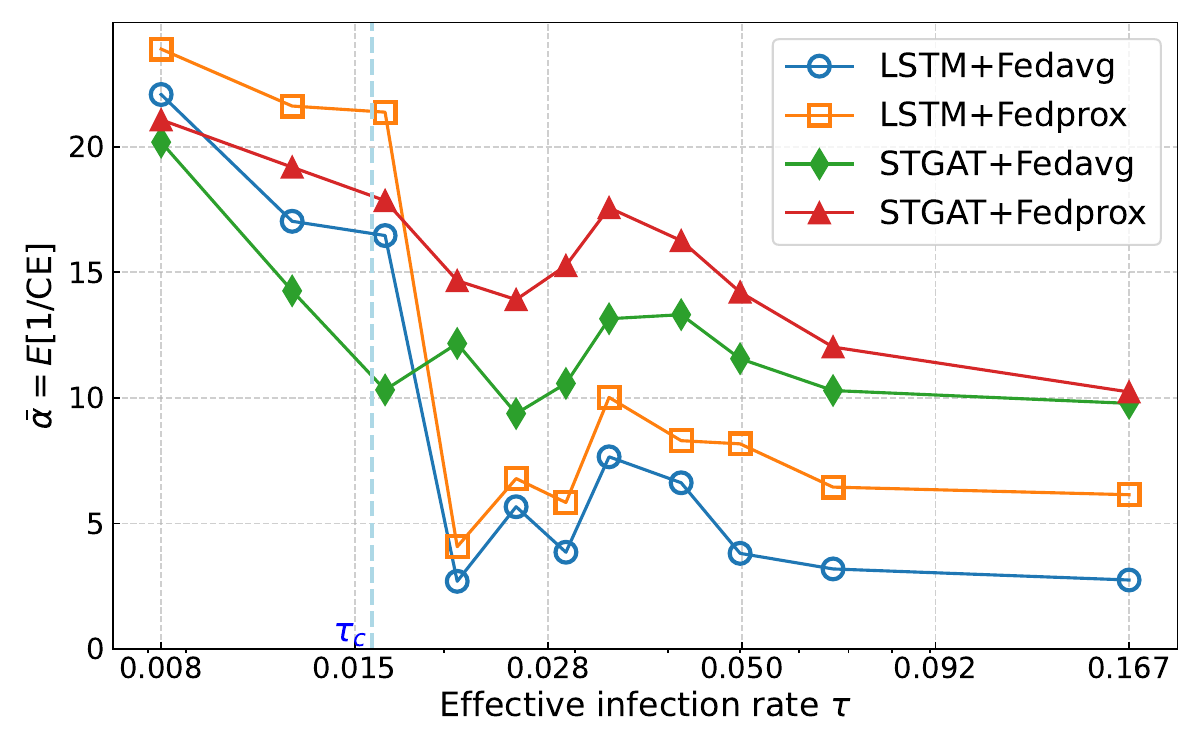}} \hfil
	\subfloat[SIS]
	{\includegraphics[width=8cm]{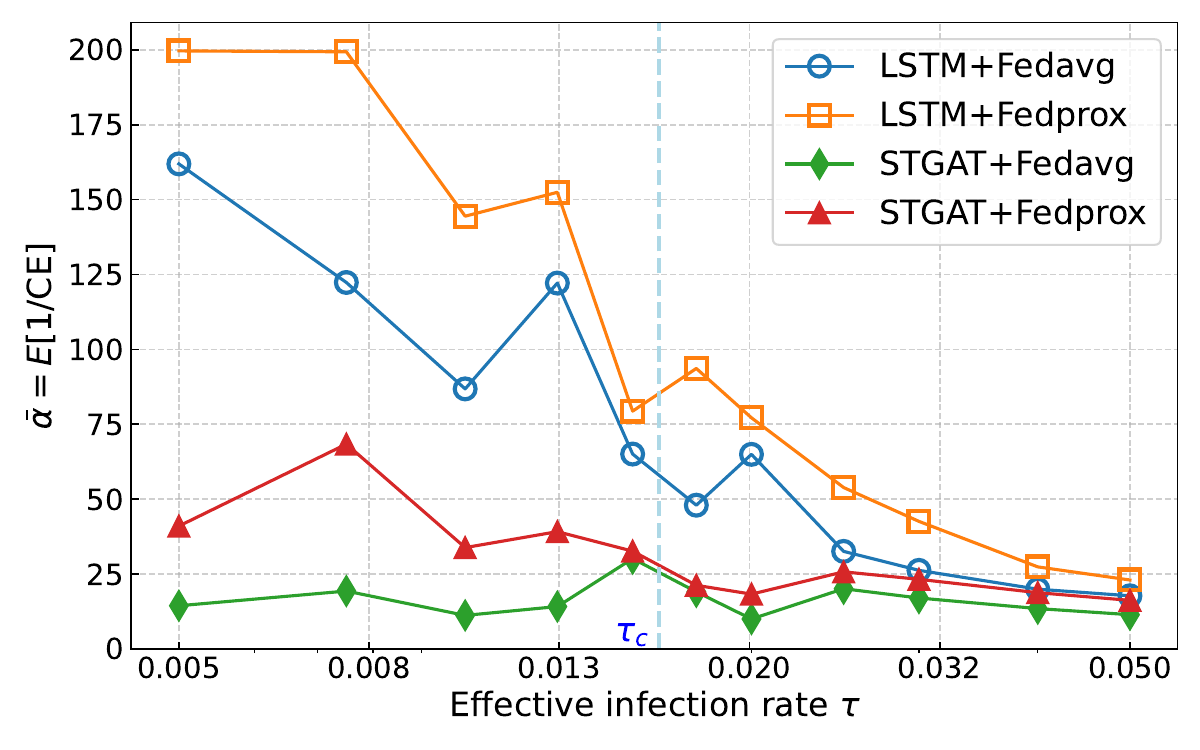}}\\
	\subfloat[SIRS]
	{\includegraphics[width=8cm]{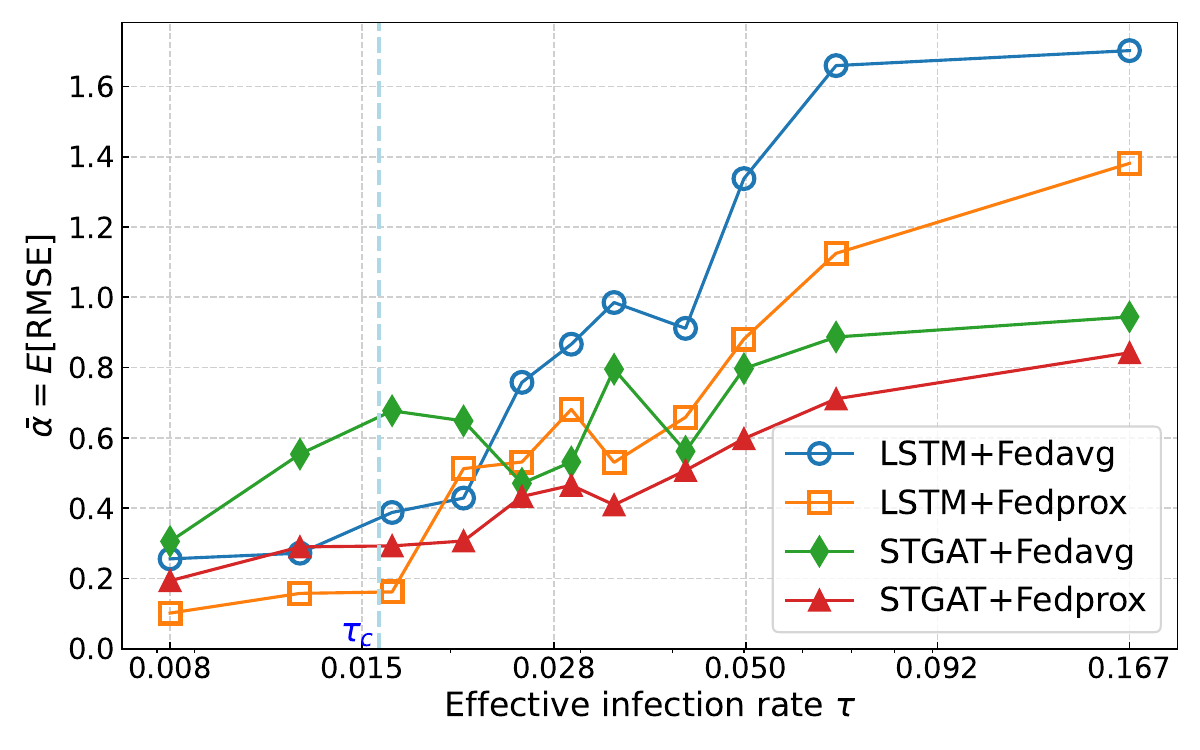}} \hfil
	\subfloat[SIS]
	{\includegraphics[width=8cm]{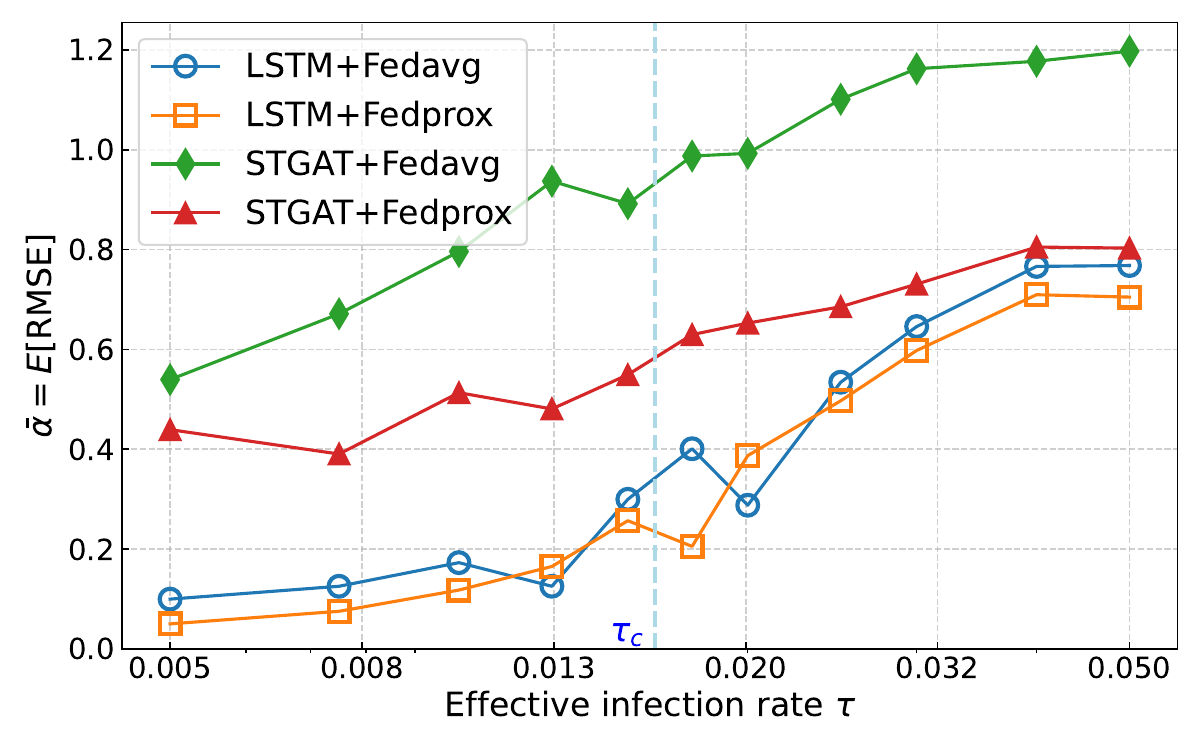}}
	\caption{Performance metrics $\bar{\alpha}=E[1/CE]$ and $\bar{\alpha}=E[RMSE]$ as a function of the effective infection rate $\tau$ for nmSIS and SIS processes under a federated learning with $M=10$ clients. The epidemic threshold $\tau_c \approx 0.015$ is marked as a vertical line. Note that a higher metric $\bar{\alpha}=E[1/CE]$ indicates better performance while a lower $\bar{\alpha}=E[RMSE]$ indicates better.}
\label{fig:infection_rate_8clients}
\end{figure}

Figure \ref{fig:infection_rate_8clients} shows that the overall prediction performance deteriorates as $\tau$ increases, albeit with minor fluctuations. This can be attributed to the dual effects of a higher effective infection rate: 
on the one hand, a larger $\tau$ introduces more dynamic variations during the early stages of the epidemic, offering richer information for model learning; on the other hand, it leads to heightened randomness in node states and increasingly complex dynamics, which complicate the prediction task.
Figure \ref{fig:infection_rate_8clients}(a)(c) shows that the performance of different FL scenarios may vary with the effective infection rate. Specifically, LSTM performs better at relatively low infection rates in SIRS where the dynamics prefer a transient decay toward extinction, while STGAT outperforms LSTM at higher infection rates. This intriguing fact indicates that the choice of FL scenario should be tailored to the characteristics of the epidemic process.

Across different scenarios, including variations in prediction models and aggregation methods, the performance metrics tend to converge at a low level as $\tau$ becomes sufficiently large. This convergence demonstrates the diminishing relative advantages of specific models or aggregation methods under the high randomness and chaotic behavior associated with large infection rates.
Our investigation underscores the intricate relationship between epidemic phase transitions and federated prediction performance. While the enhanced dynamism in epidemics can improve training efficacy, the increasing unpredictability of epidemic dynamics at higher infection rates poses significant challenges to achieving accurate predictions. 
This dilemma verifies the difficulty of providing reliable predictions in real-world scenarios characterized by rapid and widespread outbreaks.

\subsection{Sensitivity of missing reporting}
In practical scenarios, timely and comprehensive reporting of infection data is often challenging due to limitations in statistical capacity or the inability to detect infected individuals, resulting in missing or noisy data. Such inaccuracies can introduce noise into the federated learning process, potentially affecting model performance.
To investigate the sensitivity of federated epidemic prediction to missing reporting, we simulate this phenomenon by randomly selecting a fraction of clients (i.e., Client Ratio) and injecting noise into their training and validation datasets. Specifically, a proportion of the infection states is replaced with susceptible states, and this proportion is referred to as the node missing ratio.

\begin{figure}[tb]\centering 
	\subfloat[$\bar{\alpha}={E[Acc]}$ for nmSIS]
	{\includegraphics[width=8cm]{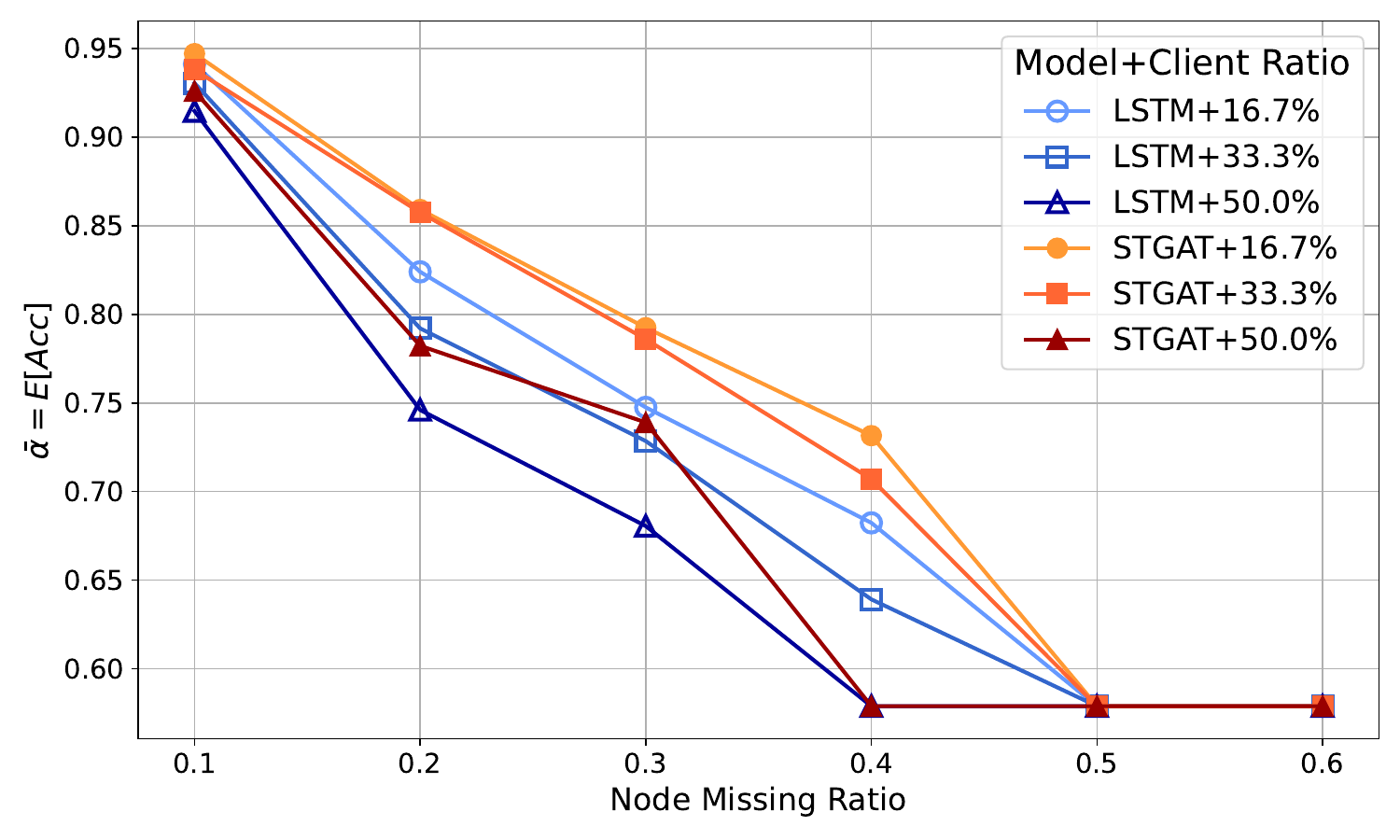}} \hfil
	\subfloat[$\bar{\alpha}={E[RMSE]}$ for nmSIS]
	{\includegraphics[width=8cm]{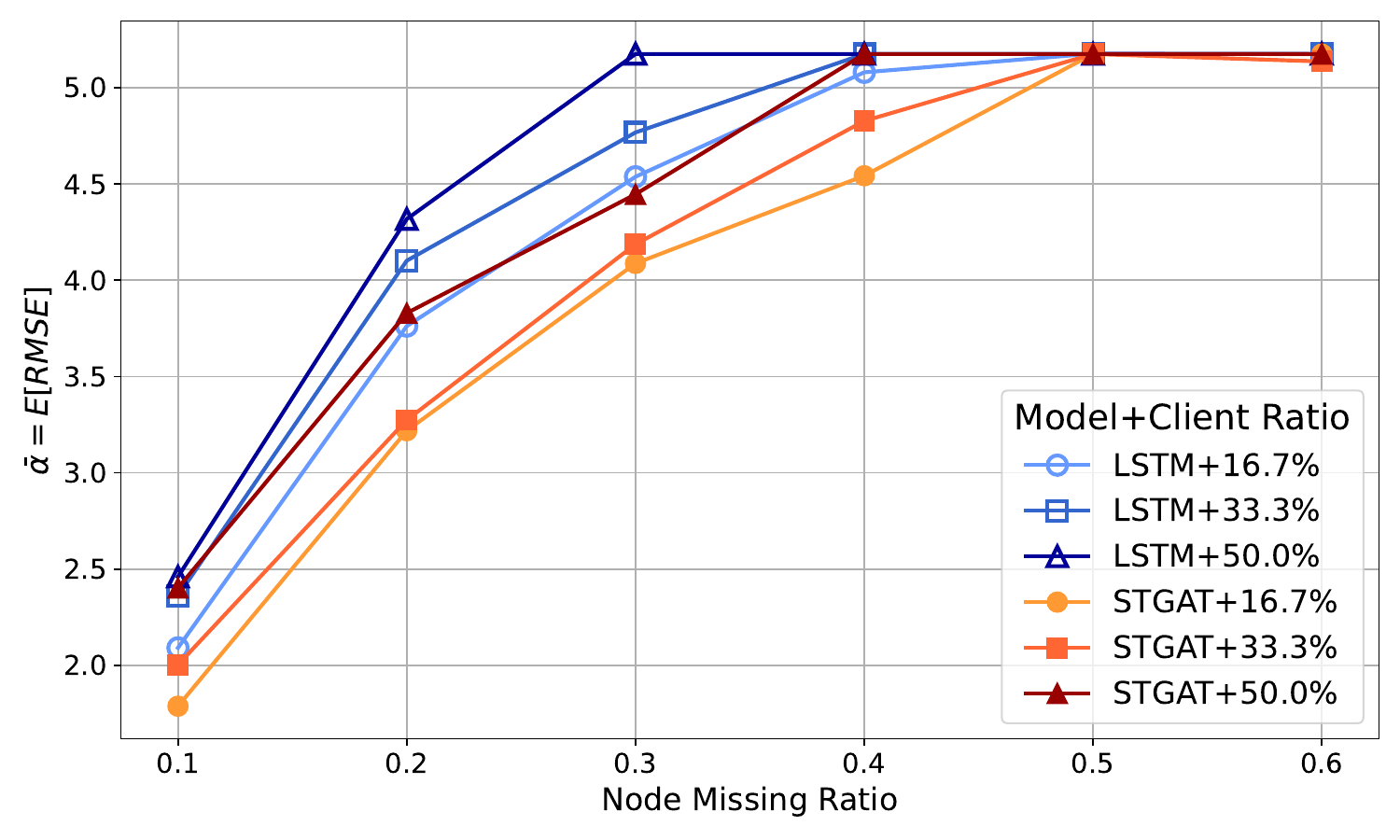}}
	\caption{Performance metric $\bar{\alpha}$ of accuracy and RMSE as a function of the node missing radio under a nmSIS federated learning of $M=6$ clients.}
	\label{fig:missing_report_nmSIS}
\end{figure}

Figure \ref{fig:missing_report_nmSIS} illustrates the metric $\bar{\alpha}$ of the accuracy and the RMSE as a function of the node missing ratio in an nmSIS federated learning scenario with $M=6$ clients. The analysis considers two prediction models under varying client ratios ($16.7\%,33.3\%,50.0\%$).
We observe that both the accuracy and the RMSE degrade with the increasing node missing ratio and the increasing fraction of delayed clients, which implies the detrimental impact of missing reporting on the model's ability to accurately capture and predict epidemic dynamics.
Consistent with the behavior of performance metrics as the number of clients increases, the metrics tend to converge at a low level when the noise becomes excessively large. This suggests that excessive missing data severely limits the learnability of meaningful patterns, leading to uniformly poor performance across different scenarios.

Our findings emphasize the critical importance of ensuring reliable, timely, and comprehensive data reporting for achieving effective federated learning in the context of epidemic prediction. 
We show the necessity to develop and adopt robust federated learning frameworks and algorithms capable of mitigating the adverse effects of noisy, incomplete, or delayed data. 
Potential solutions include incorporating noise-resilient aggregation methods, adaptive weighting of client contributions based on data quality, or integrating imputation techniques to address missing or corrupted data during training and aggregation, which may  improve the robustness and effectiveness of federated epidemic prediction in real-world.

\section{Related work}

\subsection{Epidemic prediction on networks}
Epidemic prediction has gained significant attention in recent years due to its critical role in public health decision-making and resource allocation. The primary goal is to forecast the future trajectory of disease outbreaks by leveraging historical data, environmental factors, and social interactions\cite{zino2021analysis}. 
Recent studies apply machine learning and deep learning algorithms to enhance the performance of epidemic forecasts in networked populations\cite{bamana2024systematic}. These methods excel in modeling high-dimensional, non-linear relationships and have been applied to predict disease incidence, spread patterns, and risk assessment\cite{xu2023epidemic}. Models like LSTM and GRU are utilized to capture temporal dependencies in epidemic progression \cite{shahid2020predictions}. Graph Neural Networks (GNNs) have emerged as powerful tools for leveraging the graph structure of contact networks. Methods such as Graph Convolutional Networks (GCNs), Graph Attention Networks (GATs) and Temporal Graph Networks have been used to predict disease states at individual and population levels\cite{yu2021interaction,gao2021stan,liu2024review}. 

Despite significant advancements, epidemic prediction on networks still faces several challenges. Accurate predictions rely on high-quality, granular data, which is often difficult to obtain, especially in resource-limited settings\cite{roda2020difficult}. Additionally, ensuring models generalize effectively across diverse population structures and various diseases remains a complex task\cite{desai2019real}. The ability to integrate real-time data updates into predictive frameworks is also critical for enabling timely and effective intervention planning\cite{desai2019real} \cite{alberti2020uncertainty}. Privacy protection presents another significant challenge, as the sensitive nature of health and mobility data raises concerns about data sharing and security. Striking a balance between data utility and privacy remains a pressing issue\cite{igwama2024big}. 

\subsection{Federated learning for time-series prediction}
Aimed to solve the lack of utility information scarcity and privacy preservation, federated learning focuses on effectively training global model with distributed data across numerous cross-silo clients \cite{li2020review}, allowing local users to access high-equality services without transmitting confidential information. This method is especially relevant for time-series data, given its widespread use in sensitive applications like healthcare, finance, and IoT systems\cite{imtiaz2020privacy, imteaj2022leveraging, liu2020deep}. 

Time-series data is often characterized by non-i.i.d. (non-independent and identically distributed) data distributions. Federated learning methods have developed strategies to handle such heterogeneity, which is more pronounced in time-series data due to variable sampling rates and data sparsity \cite{huang2023stochastic}. Techniques like personalized federated learning and model agnostic meta-learning extensions are increasingly applied to address these issues \cite{pei2022retracted}. Furthermore, some studies emphasize enhancing the efficiency and scalability of federated learning algorithms for processing time-series data\cite{hao2024multi}.  Another ongoing discussion focuses on how federated learning models can adapt dynamically to these changes without requiring frequent re-training or data transfer \cite{salazar2024unveiling} \cite{mavromatis2024flame}. Recent studies have also explored applications in fields such as epidemic \cite{molaei2024federated} \cite{fu2024privacy} and traffic prediction \cite{shen2024decentralized}. These applications have incorporated attention to the spatial dimension based on time-series, enabling better adaptability and improved performance. 

\section{Conclusion}
This paper presents a thorough investigation into the application of federated learning (FL) for node-level epidemic prediction on networks, highlighting its promise as a privacy-preserving and collaborative solution to address the challenges associated with data sharing in public health. 
By formulating a generalized FL framework, the research demonstrates its adaptability to various epidemic processes and its potential to capture epidemic dynamics through the traditional temporal models like LSTM and the proposed spatio-temporal models such as STGAT. These models exhibit complementary strengths, with STGAT excelling in scenarios with complex spatio-temporal dependencies and LSTM performing competitively in simpler dynamic patterns. The fact that some prediction models fail to benefit from topology information has also been observed in other fields \cite{kirschstein2024merit}.

The introduction of the efficacy energy metric provides a novel measure of system resilience, particularly under uncertain client configurations. Extensive experiments conducted on airline networks evaluate the efficacy of FL-based epidemic prediction, offering insights into the critical factors influencing performance. The results demonstrate the impact of client numbers, aggregation methods, graph partitioning strategies, and data reliability on the overall effectiveness of federated prediction. Notably, FedProx emerges as a robust choice, outperforming FedAvg in mitigating data heterogeneity and stabilizing training processes, thereby enhancing the reliability of federated epidemic prediction systems.
The findings also manifest the importance of balancing feature consistency and volume uniformity among clients in federated settings, as demonstrated by the investigation into graph partitioning. In addition, the observed degradation in efficacy performance could verify the difficulty of providing reliable predictions for a rapid epidemic spreading in real-world scenarios.
These insights establish a foundational understanding of the potential and challenges of federated learning for epidemic prediction, offering practical guidance for optimizing federated systems in epidemic management.

While the current study offers valuable insights, it also identifies several avenues for future exploration. Firstly, the analysis is limited to static networks, whereas the inclusion of temporal networks with node movement would provide a closer approximation to real-world epidemic scenarios. Secondly, the feasibility of FL in epidemic prediction highlights its potential for analyzing nonlinear complex behaviors, such as emergence, bifurcation, and stable states, even in the absence of intrinsic physical knowledge. Extending FL frameworks to other collective dynamic phenomena, such as ecological interactions \cite{gao2016universal} and transport on networks \cite{he2023heterogeneity}, represents a promising direction for future research.

\bibliographystyle{IEEEtran}
\bibliography{bibnew}

\appendix

\end{document}